\documentclass{iopart}
\usepackage[T1]{fontenc}
\usepackage{graphicx}
\usepackage{url,graphics,epsfig}
\usepackage{amssymb}

\begin{document}

\jl{2}
%
%
%
\def\etal{{\it et al~}}
%
%
%
%
%
%
\setlength{\arraycolsep}{2.5pt}             

\title[Photoionization of Be-like ions]{State-resolved valence shell photoionization of Be-like ions: experiment and theory}

\author{A M\"{u}ller$^1\footnote[2]{Corresponding author, E-mail: Alfred.Mueller@iamp.physik.uni-giessen.de}$,  S Schippers$^1$, R A Phaneuf$^2$,   A L D Kilcoyne$^3$, \\
H Br\"{a}uning$^1\footnote[1]{Present address: GSI Helmholtz Centre for Heavy Ion Research GmbH, 64291 Darmstadt, Germany  }$, A S Schlachter$^3$, M Lu$^2$, and B M McLaughlin$^{4,5}$}

\address{$^1$Institut f\"{u}r Atom- ~und Molek\"{u}lphysik,
                         Justus-Liebig-Universit\"{a}t Giessen, 35392 Giessen, Germany}

\address{$^2$Department of Physics, University of Nevada,
                          Reno, NV 89557, USA}

\address{$^3$Advanced Light Source, Lawrence Berkeley National Laboratory, Berkeley, CA 94720, USA }

\address{$^4$Centre for Theoretical Atomic, Molecular and Optical Physics (CTAMOP),
                          School of Mathematics and Physics, The David Bates Building, 7 College Park,
                          Queen's University Belfast, Belfast BT7 1NN, UK}

\address{$^5$Institute for Theoretical Atomic and Molecular Physics,
                          Harvard Smithsonian Center for Astrophysics, MS-14,
                          Cambridge, MA 02138, USA}

%
%
\begin{abstract}
High-resolution photoionization experiments were carried out using
beams of Be-like C$^{2+}$, N$^{3+}$, and O$^{4+}$ ions with roughly equal populations of the $^1$S ground-state and the $^3$P$^o$ manifold of metastable components. The energy scales of the experiments are calibrated with uncertainties of 1 to 10~meV depending on photon energy. Resolving powers beyond 20,000 were reached allowing for the separation of contributions from the individual metastable $^3$P$^o_0$, $^3$P$^o_1$, and $^3$P$^o_2$ states. The measured data compare favourably with semi-relativistic Breit-Pauli R-matrix calculations.  \\

\noindent
(Some figures in this article are in colour only in the electronic version)

\end{abstract}
%
%

\vspace{1.0cm}
\begin{flushleft}
Short title: Photoionization of Be-like ions\\
\vspace{1cm} Draft for J. Phys. B: At. Mol. \& Opt. Phys: \today
\end{flushleft}

\maketitle
%
%
%

\section{Introduction}

The most abundant elements in the universe next to hydrogen and helium are oxygen and carbon with nitrogen not far off in abundance. On earth we experience neutral C, N, and O as  main constituents of organic matter.  In most parts of the universe, however, baryonic matter is ionized and the charge states of atoms depend on the temperature of their local environment and on the degree of irradiation by energetic photons from external sources. Photoionization of ions is an important mechanism for the production of highly charged ions in cold astrophysical plasmas exposed to hot sources of radiation. Ionization in such plasmas is usually balanced by low-energy electron-ion recombination \cite{Ferland2003a}. Photoabsorption by ions (and atoms) in the interstellar medium modifies the radiation spectrum of distant objects in the universe and thus complicates the interpretation of observations of such objects. On the other hand, the observation of absorption lines of neutral or ionized gas illuminated by a continuous background electromagnetic spectrum provides detailed knowledge about the atomic abundances in the gas as well as its density and temperature. Hence, photoionization and the inverse process, photorecombination, of ions are important phenomena in astrophysical plasmas. Berylliumlike ions in particular are among the most abundant species in ionized gases owing to their relative stability as four-electron closed-shell systems. They serve as sensitive probes for the diagnostics of astrophysical and technical plasmas. Detailed knowledge of their structure and behaviour in atomic interactions, particularly in photoionization processes, is thus
essential for understanding astrophysical observations.

Besides their importance in plasma applications, Be-like ions with their two loosely bound electrons in the L-shell and two tightly bound K-shell electrons are almost perfect objects for investigating two-electron effects. Photoionization of ions in general is of basic physics interest as a fundamental process of nature. Advances in computer and parallel programming techniques have facilitated extremely accurate calculations for systems with four electrons. Thus, the investigation of photoionization of Be-like ions is an ideal testing ground for our understanding of fundamental interactions and structure in few-electron systems. As such, photoionization of Be-like ions is a subject of continuing interest. Recent publications address advances both in theoretical and experimental approaches \cite{Chu2010a,Kim2010,Simon2010}.

Theoretical work on photoionization of Be-like systems started with a calculation by Altick \cite{Altick1968} of the near-threshold photoionization cross section of the Be atom using the
configuration-interaction method. Since then photoionization of neutral Be has been treated in numerous theoretical studies. References are given in the most recent work by Hsiao \etal \cite{Hsiao2008a,Hsiao2009a}. Also ions in the Be sequence were addressed early on \cite{Flower1968}. Quantum defect theory was applied to obtain photoionization of atoms, molecules and ions \cite{Dubau1984}. Photoionization cross section calculations for C$^{2+}$ were carried out employing a close-coupling method \cite{Drew1982}. Results for the singlet ground level and, for the first time,  also for the lowest lying, metastable triplet level were presented. A substantial extension of the early theoretical studies was accomplished by the work of Tully \etal~\cite{Tully1990a} using the non-relativistic R-matrix technique to calculate photoionization cross sections in LS - coupling for the beryllium isoelectronic sequence starting with Be and reaching up to Fe$^{22+}$. The data obtained for different ions and especially for the beryllium sequence are available from the Opacity Project \cite{Seaton1994} and can be retrieved from the TOPBASE database (http://rxte.gsfc.nasa.gov/topbase/). By using more extended eigenfunction expansions Nahar and Pradhan calculated photoionization cross sections for the berylliumlike C$^{2+}$ and N$^{3+}$ ions \cite{Nahar1997}. The calculations were still performed in LS coupling and therefore could not predict the possible additional resonance structure that may arise due to intermediate coupling. In a subsequent paper Nahar provided the results of related similar calculations for O$^{4+}$ ions \cite{Nahar1998a}.

A new step forward in the theoretical treatment of photoionization of Be-like ions was made by McLaughlin by including relativistic effects in the R-matrix calculations within the confines of the semi-relativistic Breit-Pauli approximation and performing them for intermediate coupling. That work and its results for the B$^+$ and C$^{2+}$ ions were described in experimentally motivated papers by Schippers \etal \cite{Schippers2003b} and M\"{u}ller \etal \cite{Mueller2002b,Mueller2003b}, respectively. It included data not only for the $^1$S ground state of the Be-like parent ions but also for their $^3$P$^o$ metastable levels. Stimulated by the theoretical and experimental progress, Kim and Manson carried out calculations of both ground- and metastable-state photoionization of B$^+$ and C$^{2+}$ ions using a noniterative eigenchannel R-matrix method \cite{Kim2004a,Kim2004b,Kim2005a}. Kim and Kim carried out an extensive nonrelativistic calculation for the photoionization cross sections of the first excited p-states of the C$^{2+}$ ion~\cite{Kim2007a}. The computations were carried out on a fine energy mesh across all the autoionizing Rydberg series of resonances converging to Li-like C$^{3+}$ 2p, 3s, 3p and 3d thresholds.

More recently, Chu \etal studied photoionization of the beryllium isoelectronic series from neutral Be to Fe$^{+22}$ for ground $^1$S and metastable $^3$P$^o$ initial states \cite{Chu2009}. Both nonrelativistic LS-coupling R-matrix and relativistic Breit-Pauli R-matrix methods were used to calculate the cross sections in the photon-energy range between the first ionization threshold and the 1s$^2$\,4f$_{7/2}$ threshold for each series member. The total cross sections compare well with available experimental results. The importance of relativistic effects was demonstrated by the comparison between the LS and the Breit-Pauli results. The same group of authors has just published R-matrix  results on photoionization of Be-like ions including processes of ionization plus excitation of the target ion \cite{Chu2010a}. Another approach to photoionization was also presented recently. Hsiao \etal applied the multiconfiguration relativistic random-phase approximation theory to the valence-shell photoionization of ions along the Be isoelectronic sequence \cite{Hsiao2009a}. The most recent theoretical work addresses total and partial cross sections for photoionization of the Be-like O$^{4+}$ ion. The calculations were based on a non-iterative variational R-matrix method combined with multichannel quantum-defect theory for the ground 2s$^2$ $^1$S and excited 2s2p $^{1,3}$P states. The photon energy region from the first threshold up to the O$^{5+}$ 3d threshold was covered.

Experimental work on photoionization of Be-like systems started with the observation of far-ultraviolet absorption spectra of beryllium vapor produced in a vacuum spark arrangement backlit from a second Be vacuum spark \cite{Mehlman-Balloffet1969}. Photoionization of neutral Be atoms was studied in detail by exposure of Be metal vapor to synchrotron radiation \cite{Wehlitz2005a,Olalde-Velasco2007}. The absorption spectrum of Be-like B$^+$ was investigated by using two laser produced plasmas \cite{Jannitti1986}; one generating the B$^+$ absorbing medium and the other the background continuum radiation. Series of lines arising from the  $^1$S ground and the $^3$P$^o$ metastable levels were observed followed by absorption continua at higher energies. Photoionization cross sections were derived and autoionization resonances  observed. The associated resonance profiles were fitted with a parametric formula for series of autoionized lines \cite{Jannitti1986} developed on the basis of quantum defect theory \cite{Dubau1984}.

With the development of photon-ion merged beams experiments located at synchrotron radiation facilities systematic measurements of photoionization cross sections along isoelectronic and isonuclear sequences became possible. The techniques and applications of merged beams have been reviewed by West \cite{West2001a} and more recently by Kjeldsen \cite{Kjeldsen2006a}. The berylliumlike sequence was addressed previously in a number of merged beams experiments providing absolute cross sections up to the (2pnl) resonance-series limits for photoionization of B$^{+}$ ions  with an energy resolution of 25~meV at around 30~eV photon energy \cite{Schippers2003b}, for photoionization of C$^{2+}$ with an energy resolution of 30~meV at around 50~eV photon energy \cite{Mueller2002b}, for photoionization of N$^{3+}$ with a quoted energy resolution of 230~meV at around 70~eV photon energy \cite{Bizau2005a}, and for photoionization of O$^{4+}$ with an underestimated energy resolution of 250~meV at around 110~eV photon energy \cite{Bizau2005a,Champeaux2003a}. For selected resonances in the photoionization of C$^{2+}$ ions measurements with energy spreads as low as 7.5~meV have been reported \cite{Mueller2003b}. An investigation of state-selective photoexcitation of autoionizing levels of  C$^{2+}$, N$^{3+}$, and O$^{4+}$ ions \cite{Mueller2007b} was launched a few years later to test theory at an unprecedented level of detail. Very recently, photoionization of highly charged ions trapped in an Electron-Beam-Ion-Trap (EBIT) has been demonstrated and data for photoionization of Be-like N$^{3+}$ ions were presented \cite{Simon2010}. The autoionizing 2p5p resonances arising from excitation of metastable N$^{3+}$(1s$^2$\,2s2p $^3$P$^o$) ions by photons with energies between 69.7 eV and 70.1 eV were measured with an energy spread of about 35~meV.

All the Be-like ions feature long lived (2s2p $^3$P$^o$) states which can be easily populated in hot ion sources. Lifetimes range between 470 $\mu$s for O$^{4+}$(2s2p $^3$P$^o_1$)
\cite{Froese-Fischer2004} to essentially $\infty$ for the $^3$P$^o_0$ states. Only in the presence of nuclear magnetic moments, i.e., for isotopes with un-paired protons or neutrons, the lifetime of the $^3$P$^o_0$ state is somewhat shortened by the effect of hyperfine quenching~\cite{Schippers2007a}.  As a result, practically none of the possible beam components can decay during the few-$\mu$s flight times of the ions between the source and the photon-ion interaction region in a merged-beams experiment. Even long storage times in storage rings or ion traps  do not guarantee complete elimination of all metastable states in an ensemble of Be-like ions with low to moderately high atomic numbers Z. Measurements on Be-like ions thus provide information on photoionization of both the ground-state and long-lived metastable states. The fractional populations of the individual states are not easily accessible and unknown fractions of different states in the primary ion beam result in uncertainties of the measured cross section data. As a result, the demands on theory are relaxed. This disadvantage can be compensated for by resolving individual contributions to the apparent photoionization cross section from the different excited components of the primary ion beam. The
high-resolution high-flux photon beams of the Advanced Light
Source (ALS) in Berkeley can be exploited to accomplish such separation of
contributions of the $^3$P$^o_0$, $^3$P$^o_1$,
and $^3$P$^o_2$ parent ion states. On the basis of high-precision experimental data on Be-like C$^{2+}$, N$^{3+}$, and O$^{4+}$ ions, theories such as the present state-of-the-art Breit-Pauli R-matrix approach to photoionization of ions can be tested at a new unprecedented level of detail.

The layout of this paper is as follows. Section 2 presents a brief outline of the theoretical work.
Section 3 details the experimental procedure used. Section 4 with subsections for the 3 investigated Be-like ions presents an extensive discussion of the
results obtained from both the experimental and the theoretical methods employed.
Finally in section 5 conclusions are drawn from the present investigation.

\section{Theory}\label{sec:theory}

The processes investigated in the present study can be described in LSJ terms as
\begin{equation}\label{eq:gsprocess}
h\nu + 1{\rm s}^2 2{\rm s}^2\,^1{\rm S}_0^e \to \{1{\rm s}^2 {\rm nl} + {\rm e}^-({\rm kl'})\}\,^1{\rm P}_1^o
\end{equation}
for photoionization of ground-state Be-like ions and
\begin{equation}\label{eq:msprocess}
h\nu + 1{\rm s}^2 2{\rm s}2{\rm p}\,^3{\rm P}_{0,1,2}^o \to \{1{\rm s}^2 {\rm nl} + {\rm e}^-({\rm kl'})\}\,^3{\rm S}_1^e,\, ^3{\rm P}_{0,1,2}^e,\, ^3{\rm D}_{1,2,3}^e
\end{equation}
for photoionization of metastable Be-like ions. R-matrix calculations were carried out to determine photoionization cross sections for the $^1$S ground and the $^3$P$^o$ metastable ($N+1$ -- electron) initial states of the Be-like ions C$^{2+}$, N$^{3+}$, and O$^{4+}$. Here, $N$ is the number of electrons in the 'core' of the ion ($N=3$ in the case of Be-like ions) while 'the active electron' is counted as number $N+1$. For generating the $N+1$ -- electron wavefunction of the Be-like ions the present R-matrix work started from the construction of nine Li-like LS states; 1s$^2$2s~$^2$S, 1s$^2$2p~$^2$P$^o$, 1s$^2$3s~$^2$S, 1s$^2$3p~$^2$P$^o$, 1s$^2$3d~$^2$D,  1s$^2$4s~$^2$S, 1s$^2$4p~$^2$P$^o$, 1s$^2$4d~$^2$D and 1s$^2$4f~$^2$F$^o$ to represent the possible final states of the photoionized ions C$^{3+}$, N$^{4+}$, and O$^{5+}$, respectively. The final state of the investigated photoionization processes is an $N$-electron state (termed the target state for historical reasons). The target ion core is retained in the close-coupling expansions for both the $^1$S ground and the $^3$P$^o$ metastable initial states of the Be-like ions. In the structure calculations for the Li-like target ions, all physical orbitals were included up to n=4 in the configuration interaction wavefunctions used to describe the states. The
Hartree-Fock 1s and 2s orbitals of Clementi and Roetti \cite{Clementi1974} were used, with the additional n=3 and n=4 orbitals determined by energy optimization on the appropriate spectroscopic state with the structure code CIV3 of Hibbert \cite{Hibbert1975}. All of the nine states of each Li-like core were represented by multi-configuration interaction wavefunctions. The Breit-Pauli R-matrix approach was utilized to calculate the fifteen C$^{3+}$(LSJ) (N$^{4+}$(LSJ), O$^{5+}$(LSJ)) target ion state energies which arise from the nine LS states, and the photoionization cross sections were determined for both the $^1$S ground and $^3$P$^o$ metastable initial states.

PI cross-section  calculations for  Be-like ions were performed in intermediate coupling using the semi-relativistic Breit-Pauli approximation which allows for relativistic effects to be included   within the confines of the R-matrix approach~\cite{Burke1993,Berrington1995,Robicheaux1995}. The scattering wavefunctions were generated by allowing all possible three electron promotions out of the base 1s$^2$2s$^2$ configuration of each Be-like ion into the orbital set employed. All the scattering calculations were performed with twenty continuum functions and a boundary radius of 15.8 Bohr radii for the C$^{2+}$ ion, 12.8 Bohr radii for the N$^{3+}$ ion, and 10.8  Bohr radii for the O$^{4+}$ ion. For the $^1$S ground state  dipole selection rules require only the following transition: $ 0^e\rightarrow 1^o$, whereas for the
$^3$P$^o$ metastable state one needs all the dipole transitions: $2^o\rightarrow 3^e,2^e,1^e$, {\,\,} $1^o\rightarrow 2^e,1^e,0^e$, and $0^o\rightarrow 1^e$ to be calculated.
The Hamiltonian matrices for the $ 2^o$, $1^o$, $ 0^o$, $ 3^e$, $ 2^e$, $ 1^e$ and $0^e$ symmetries were calculated, where the entire range of LS matrices that contribute to these J$\pi$ symmetries are used. Effects of radiation damping~\cite{Robicheaux1995} were included.

For the $^1$S ground state and $^3$P$^o$ metastable initial
states the outer region electron-ion collision problem was solved using a suitably chosen fine energy mesh of 13.6~$\mu$eV = $10^{-6}$ Rydbergs  between the thresholds in order to fully resolve the fine resonance structure in the respective photoionization cross sections. The energy mesh of the present calculations substantially differs from that of our previous theoretical treatment on photoionization of C$^{2+}$ ions \cite{Mueller2002b}. Due to the large calculational effort required when using a fine mesh, the  energy steps were chosen to be 1.36 meV in that previous investigation, i.e. they were a factor 100 more coarse than in the present study. The effect of the size of the energy mesh is discussed in the results section of this paper.

The present state-of-the-art calculations were augmented with calculations employing the online version of the Los Alamos National Laboratory (LANL) Atomic Physics Codes package \cite{LANL}. This easy-to-use package is based on Cowan's Hartree-Fock atomic structure theory~\cite{Cowan1981} and makes use of the \underline{C}owan \underline{AT}omic \underline{S}tructure code CATS. It can provide oscillator strengths for specific transitions from a state $i$ to a state $f$ of an atom in a given charge state. It also calculates the associated transition energies although in the online version  not nearly with the accuracy of the Breit-Pauli R-matrix theory. In combination with the state-of-the-art theory and the high resolution experiments reported here, the online CATS package can be used to identify individual contributions to the photoionization cross sections studied.

\section{Experiment}\label{sec:exp}

The experiment was performed at the ion-photon-beam (IPB) end-station of the undulator beamline 10.0.1 at the ALS. A detailed description of the experimental setup has been provided by Covington \etal \cite{Covington2002a}. For the present study on Be-like C$^{2+}$, N$^{3+}$, and O$^{4+}$ ions similar procedures were utilized as in our earlier photoionization cross section  measurements for Be-like boron  and carbon ions \cite{Schippers2003b,Mueller2002b,Mueller2003b}. The berylliumlike ions were generated inside a compact all--permanent-magnet electron cyclotron-resonance (ECR) ion source \cite{Broetz2001}. Collimated ion-beam currents of
typically 30 to 100~nA  were extracted by putting the ion source on a
positive potential fixed between +5 and +6.5~kV. The ions were then passed through a bending dipole magnet selecting the desired ratio of charge to mass.  The mass- and charge-selected ion beam was centered onto the counterpropagating photon beam by applying appropriate voltages to several electrostatic ion beam steering and focusing devices. Downstream of  the
interaction region, the ion beam was deflected out of the photon beam direction by a second
dipole magnet that also separated the ionized C$^{3+}$ (N$^{4+}$, O$^{5+}$) product ions from the respective  C$^{2+}$, (N$^{3+}$, O$^{4+}$) parent ions. The product ions were counted with nearly 100\% efficiency using suitable single-particle detection \cite{Fricke1980,Rinn1982}, and the parent ion current was monitored for normalization. The measured product count rate $R$ was only partly due to photoionization events. It also contained identical ions produced by electron removal collisions of the parent ions with residual gas molecules and surfaces. This background was determined by mechanically chopping the photon beam.

In the present experiments relative energy-scan measurements were
carried out by stepping the photon energy through a preset range of values at different photon beam energy resolutions. The desired experimental energy spread was preselected by adjusting monochromator settings of the beamline  accordingly. It is not possible, however, to choose a well defined resolution by just setting the monochromator slits. The real resolution has to be determined from the measured results after the experiment. The reason for this is in the slit geometry and the small size of the photon beam at the entrance slit of the monochromator in the beamline used.
Measurements were pushed to the practical limits of resolution
which were mainly determined by limitations in the
signal-to-background ratios acceptable for experiments and by the photon beamtime available, since finer energy steps are needed to resolve features at higher resolution. Resolving powers $E/\Delta E$ of more than 20,000 were reached. The scan data were normalized to absolute measurements available from
previous work \cite{Mueller2002b,Bizau2005a}. The total
systematic uncertainty of the cross sections thus determined is
about $\pm$20~\%. Additional uncertainty arises from the presence
of unknown fractions of metastable and ground-state ions in the
parent beams.

The energy scales of the experiments are calibrated
with uncertainties of 1 to 10~meV depending on photon energy.
The calibration was accomplished by carrying out separate photoionization measurements
with neutral He, Ar (L-shell resonances in 2$^{nd}$ order), and Kr gas and by comparing the results with the well known resonance features \cite{Domke1996b,King1977} at
energies between about 60~eV and 120 eV. Calibrated monochromator settings from these ranges were linearly interpolated to obtain the scaling factors for measured energies in the present range of interest. Since the parent ions are in motion, a Doppler correction has to be carried out transforming the nominal laboratory energies to the center of mass frame of the ions before applying the calibration factor. With all this carefully included, we estimate an absolute uncertainty of at most $\pm 10$~meV for the energy scale of the present measurements, at some energies close to the calibration energies the absolute uncertainties are as small as $\pm 1$~meV. Since the precision of peak position determinations is of the order of only 0.1~meV, the possible calibration error almost exclusively determines the absolute uncertainties of the resonance energies. Relative uncertainties of resonance energies within the range of one scan measurement are very much smaller than $\pm 10$~meV. Especially in scan ranges of only about 2 to 3 hundred meV the relative uncertainties of peak positions are only about 0.1~meV.

\section{Results and Discussion}\label{sec:res}

The results of the present work are described in detail in the following subsections for each of the Be-like ions investigated. Theoretical cross sections were calculated separately for each of the initial  (1s$^2$\,2s$^2$~$^1$S$_0$), (1s$^2$\,2s2p~$^3$P$^o_0$), (1s$^2$\,2s2p~$^3$P$^o_1$) and (1s$^2$\,2s2p~$^3$P$^o_2$) states of the Be-like C$^{2+}$, N$^{3+}$, and O$^{4+}$ parent ions. The calculated unconvoluted cross sections show very rich details with numerous resonances of very different widths. In order to assess the strengths contained in these features it is convenient to convolute the calculated cross sections with a gaussian distribution of defined width. The convoluted spectra provide an immediate overview over the relative importance of the different resonance contributions. For comparison with experiments such convolution has to be carried out anyway  in order to simulate the limited experimental energy resolution.

In the experiments all the above initial states were present in the parent ion beams used for the measurements. The ECR ion source has to provide electron energies sufficiently high to produce the desired multiply charged ions. Hence, population of the metastable $^3$P$^o$ states with excitation energies of 6.50~eV for  C$^{2+}$, 8.34~eV for N$^{3+}$, and 10.19~eV for O$^{4+}$ ions \cite{NIST2004} was unavoidable. Moreover, the lifetimes of the $^3$P$^o$ states of the investigated ions are at least about 470~$\mu$s (for O$^{4+}$(1s$^2$\,2s2p~$^3$P$^o_1$)~\cite{Froese-Fischer2004} while the time of flight between the ion source and the photon-ion interaction region is at most 8~$\mu$s. Therefore ions in these metastable states constituted part of the primary ion beam.
Comparison of the theoretical cross sections with the measured (apparent) cross sections  indicates a fractional content $f_{\rm m}$ of metastable $^3$P$^o$ states in the parent ion beams with $f_{\rm m} = (0.50 \pm 0.10)$ while the fraction of ground state $^1$S ions is $f_{\rm g} = (1-f_{\rm m})$. The individual numbers found in each experiment will be discussed in the appropriate subsections.

The energy spacings between the 3 individual fine-structure states within the $^3$P$^o$ manifolds are very small as compared to the excitation energies. The splitting between the energetically lowest state, $^3$P$^o_0$, and the highest, $^3$P$^o_2$, is only 9.9~meV for C$^{2+}$, 25.7~meV for N$^{3+}$, and 54.9~meV for O$^{4+}$. The ratio of energy splitting versus excitation energy is the largest for the O$^{4+}$ ions where it amounts to not more than 0.0054. Statistical population of the fine structure states therefore appears to be a good assumption, i.e., 1/9 of $f_{\rm m}$ is in the $^3$P$^o_0$ state, 3/9 are in the $^3$P$^o_1$ state, and 5/9 of $f_{\rm m}$ are in the $^3$P$^o_2$ state. The present comparisons of theory and experiment fully support this assumption. In our previous work on photoionization of the B$^+$ ion \cite{Schippers2003b} the assumption of an overall population of metastable states was $f_{\rm m} = 0.29$ and all of that in the $^3$P$^o_1$ state gave the best result for a fit of the experiment with theory using the metastable fractions as fit parameters. The present analysis casts some doubt on those fractions. Our first study on photoionization of Be-like ions was carried out with C$^{2+}$ ions~\cite{Mueller2002b,Mueller2003b}. The calculations performed in that work were similar in nature to the present theory, but were carried out with an energy mesh of 1.36 meV, compared to a mesh of 13.6~$\mu$eV for the present calculations. It was discovered that such a finer mesh for the theory was needed to properly characterize the narrower resonances in Be-like ions, as illustrated in subsection 4.1.  Basically, the theoretical results with an energy step width of 13.6~$\mu$eV can be regarded as new results and the previously derived metastable fractions have to be revised on the basis of the present new findings.

\subsection{Photoionization of C\,$^{2+}$}\label{sec:C2}

%
%
%

\begin{figure}
\begin{center}
\includegraphics[width=\textwidth]{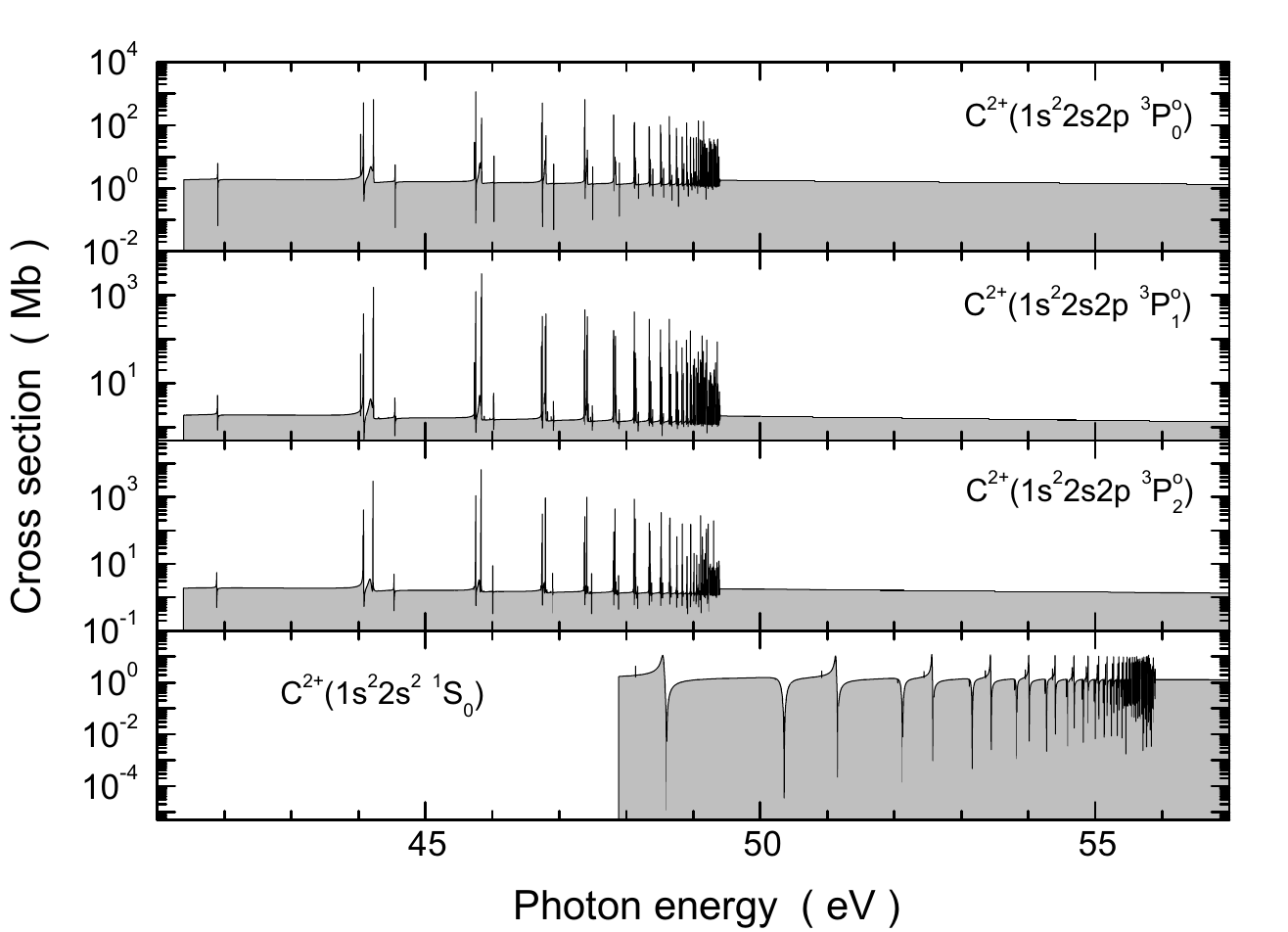}
\caption{\label{fig:C2theoryoverview} Results of the present Breit-Pauli R-matrix calculations for photoionization of C$^{2+}$ ions in the 4 different (1s$^2$\,2s$^2$~$^1$S$_0$), (1s$^2$\,2s2p~$^3$P$^o_0$), (1s$^2$\,2s2p~$^3$P$^o_1$) and (1s$^2$\,2s2p~$^3$P$^o_2$) states at an energy step width of 13.6~$\mu$eV.}
\end{center}
\end{figure}

Figure~\ref{fig:C2theoryoverview} shows the results of the present Breit-Pauli R-matrix calculations for photoionization of C$^{2+}$ ions in the 4 different (1s$^2$\,2s$^2$~$^1$S$_0$), (1s$^2$\,2s2p~$^3$P$^o_0$), (1s$^2$\,2s2p~$^3$P$^o_1$) and (1s$^2$\,2s2p~$^3$P$^o_2$) states carried out at a step width of 13.6~$\mu$eV. The cross sections are on logarithmic scales. The first question to be addressed is, whether the very fine small-width resonances that may have been missed in the previous 1.36-meV calculation~\cite{Mueller2002b,Mueller2003b} add substantial resonance strength to the photoionization spectrum. It should be noted that the continuous part of the cross section describing the direct single-ionization process is not influenced by the step width of the calculations. Integrating the two sets of spectra, calculated at step widths 1.36~meV and 13.6~$\mu$eV, over the investigated energy range results in only a small difference. The integral over the (1s$^2$\,2s$^2$~$^1$S$_0$) photoionization spectrum in the energy range from threshold to the series limit of the (1s$^2$\,2p\,nl~$^1$P) resonances gives 11.987~Mb\,eV at 1.36~meV step width and 12.001~Mb\,eV at 13.6~$\mu$eV, i.e. the strengths contained in the two spectra are practically identical. The largest difference in strengths for the older and the present calculations is found for the (1s$^2$\,2s2p~$^3$P$^o_0$) photoionization spectrum. Integrating from threshold to the series limit of the (1s$^2$\,2pnp~$^3$S,$^3$P,$^3$D) resonances gives 16.145~Mb\,eV at 1.36~meV step width and 16.724~Mb\,eV at 13.6~$\mu$eV, i.e. a difference of only about 3.5~\%.

\begin{figure}
\begin{center}
\includegraphics[width=\textwidth]{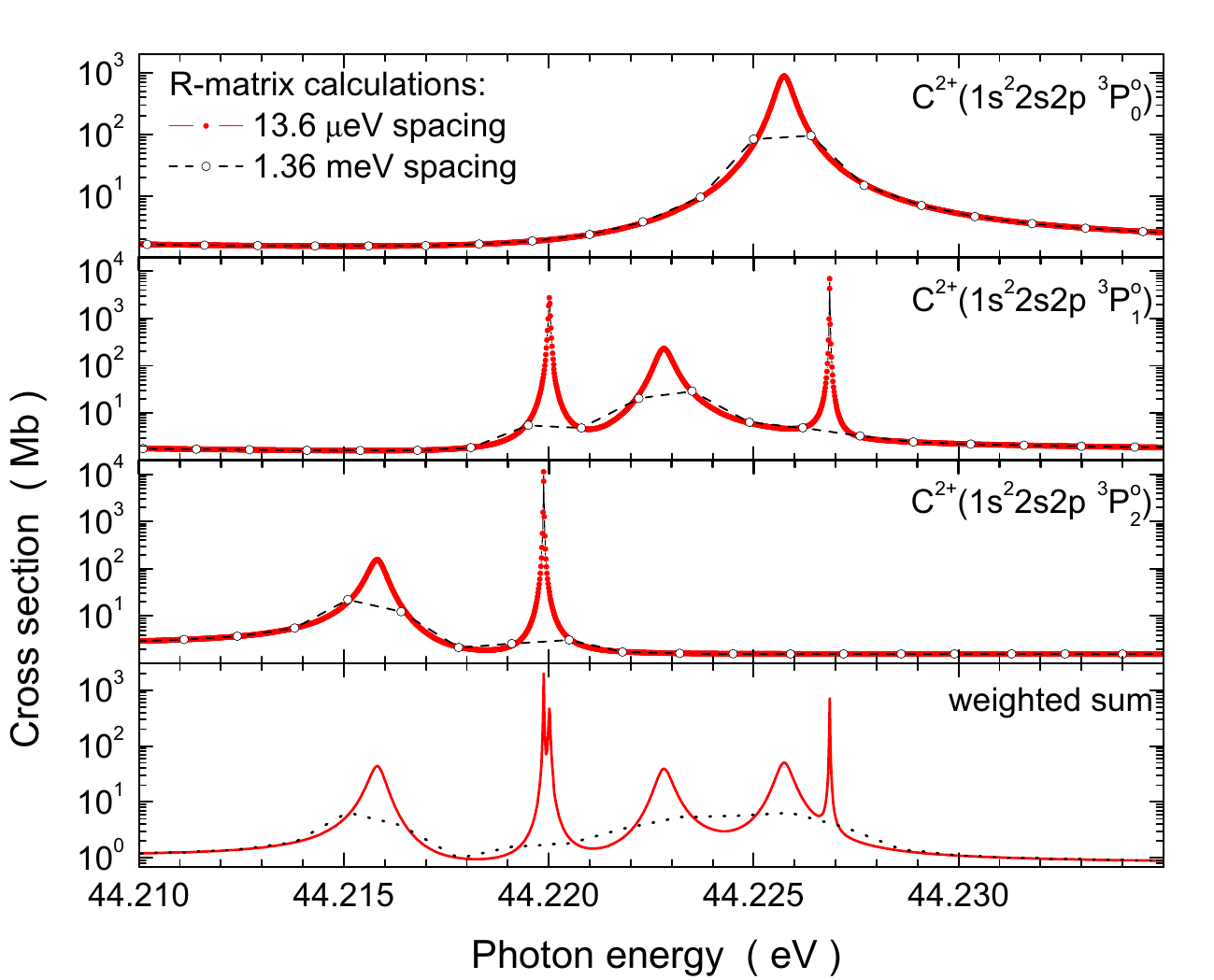}
\caption{\label{fig:C2theoryzoom} (Colour online) Results of the R-matrix calculations from figure~\ref{fig:C2theoryoverview} in a narrow energy range where only the $^3$P$^o$ states can contribute to photoionization. The present calculations with a 13.6~$\mu$eV spacing are represented by solid (red) dots connected by straight solid black lines. The previous calculations with 1.36~meV resolution \cite{Mueller2002b} are shown by the open circles connected by straight dashed lines. The lowest panel simulates experiments with infinite resolving power using a primary C$^{2+}$ ion beam with a metastable fraction $f_{\rm m} = 0.5$ and statistical population of the $^3$P$^o$ fine structure components. The solid (red) line results from the present calculations while the dotted line represents the previous theoretical data. Note the logarithmic cross section scale.
}
\end{center}
\end{figure}

While the overall picture is not changed by calculations on a narrower energy grid clear differences become obvious when zooming into the calculated spectra. An example is provided in figure~\ref{fig:C2theoryzoom} where the results of the two sets of calculations are shown in an energy range from 44.210~eV to 44.235~eV covering a span of only 15~meV. At the energies where the cross sections were calculated at 1.36~meV step width the previous and the present results agree. Linear interpolation between the calculated points, however, chops off excursions of the cross section function in narrow energy ranges. The present calculations show that some of the narrow resonances reach peak values of more than ten thousand Mb.  Again, one has to ask whether these excursions make a significant difference in experimental measurements with realistic conditions. A good way to illustrate the relative strengths contained in resonance features is to convolute the data with a gaussian function of a given width greater than the natural width of the broadest resonance in the spectrum as already discussed above. Such convolution can also be used to simulate an experiment with finite resolving power.

\begin{figure}
\begin{center}
\includegraphics[width=0.8\textwidth]{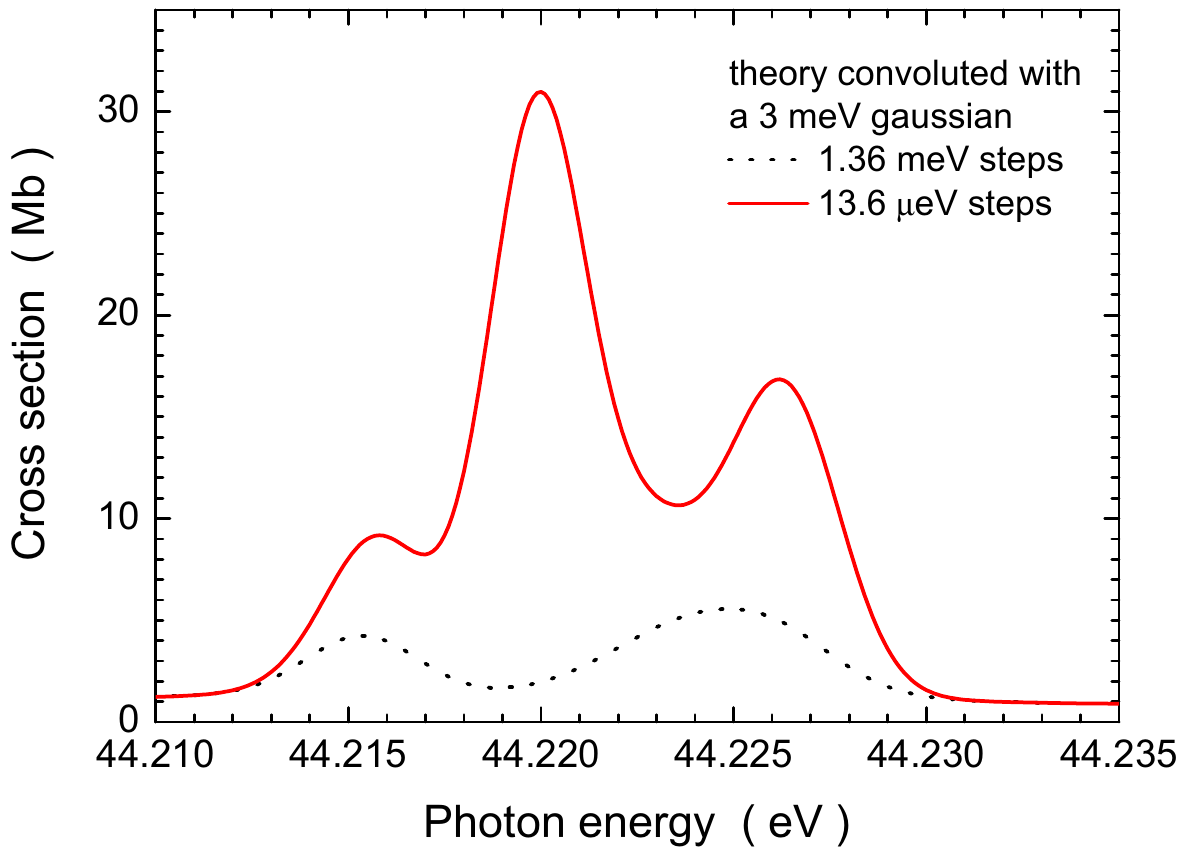}
\caption{\label{fig:C2theoryzoom3meV} (Colour online) Simulated results of an experiment on photoionization of C$^{2+}$ ions with 3~meV resolution in a narrow energy range. The primary ion components are the same as those assumed for the lowest panel of figure~\ref{fig:C2theoryzoom}. The dotted line results from the previous calculations \cite{Mueller2002b} with 1.36~meV spacing, the solid (red) line results from the present theoretical data calculated at 13.6~$\mu$eV resolution.
}
\end{center}
\end{figure}

The lowest panel in figure~\ref{fig:C2theoryzoom} shows a simulated experimental result that would be obtained at infinite resolving power using a primary C$^{2+}$ ion beam with a metastable fraction $f_{\rm m} = 0.5$ and statistical population of the $^3$P$^o$ fine structure components. The result of the convolution of this spectrum with a gaussian of 3~meV full-width-at-half-maximum (FWHM) is shown in figure~\ref{fig:C2theoryzoom3meV}. The difference between the two sets of theoretical data is striking. For testing theory at this level, however, experimental resolving powers well over 10,000 are needed.

\begin{figure}
\begin{center}
\includegraphics[width=\textwidth]{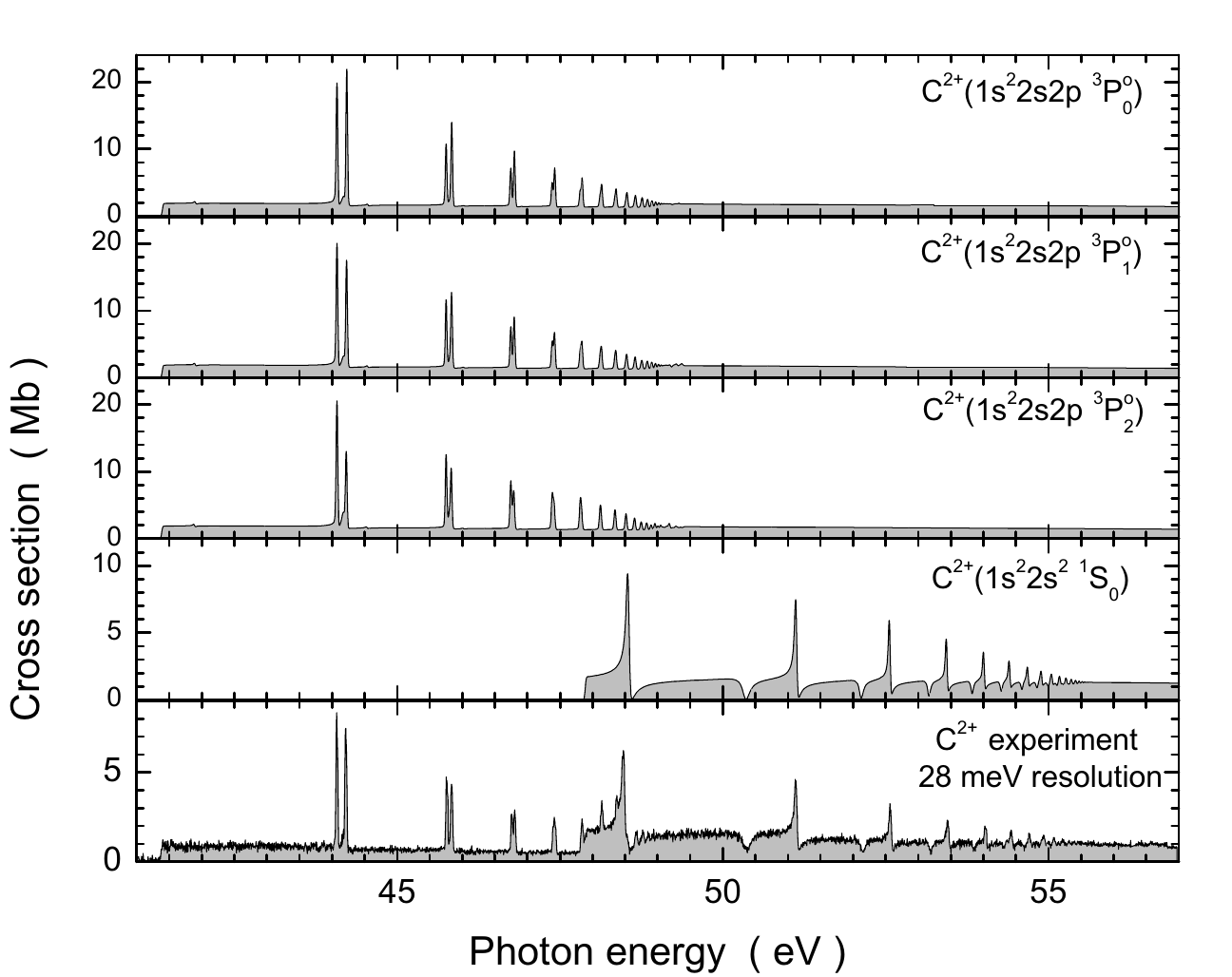}
\caption{\label{fig:C2overviewcomparison}  Comparison of the present calculations  with previous experimental data for photoionization of C$^{2+}$ ions \cite{Mueller2002b}. The theoretical data of figure~\ref{fig:C2theoryoverview} were convoluted with a 28~meV FWHM gaussian to simulate the experimental energy resolution and are shown separately for the different initial states relevant to the experiment. The lowest panel presents the measured data (solid line).
}
\end{center}
\end{figure}

\begin{figure}
\begin{center}
\includegraphics[width=\textwidth]{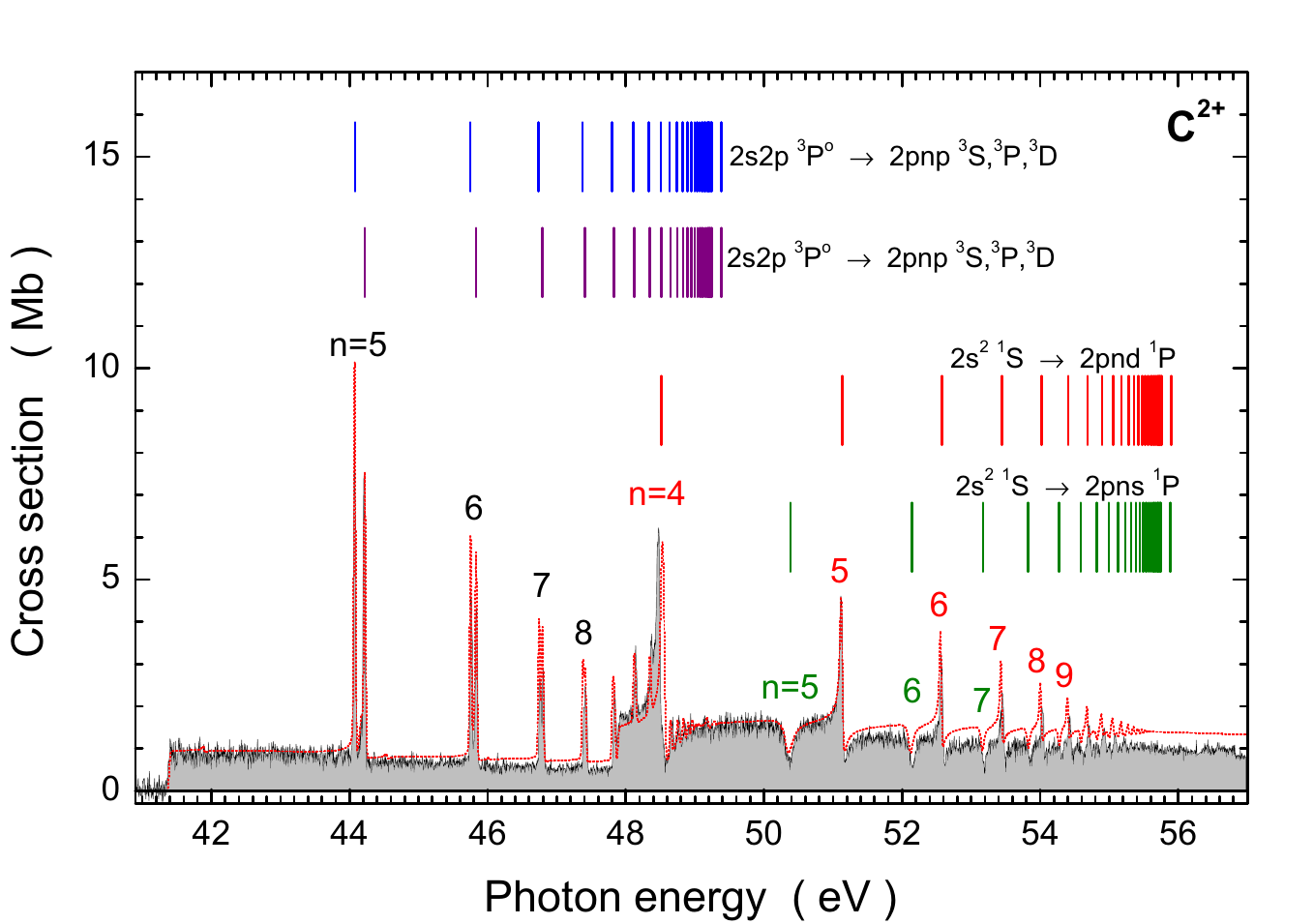}
\caption{\label{fig:C2overview} (Colour online)
Comparison of the experimental overview of C$^{2+}$ photoionization~\cite{Mueller2002b} with a model spectrum obtained from the present Breit-Pauli R-matrix calculations. The theoretical spectrum was obtained by assuming a metastable fraction $f_{\rm m}=0.5$ in the parent ion beam and statistical population of the (1s$^2$\,2s2p~$^3$P$^o_0$), (1s$^2$\,2s2p~$^3$P$^o_1$) and (1s$^2$\,2s2p~$^3$P$^o_2$) metastable states. The experimental cross section is displayed as a solid line with light gray shading. The theoretical model spectrum is represented by the dotted (red) curve. The series of vertical lines indicate the Rydberg sequences of different groups of resonances excited by the photons as the photon energy increases. The (coloured) numbers on the resonances mark the principal quantum numbers of the associated Rydberg series.
}
\end{center}
\end{figure}

The new calculations are shown in figure~\ref{fig:C2overviewcomparison} along with the published experimental overview spectrum for photoionization of the C$^{2+}$ ion \cite{Mueller2002b} measured at resolving power smaller than 2000. In the previous analysis an experimental energy spread of 30~meV was inferred. In the present re-analysis on the basis of the new calculations the experimental energy spread was found to be slightly smaller, namely 28~meV. Thus, the new theoretical data for the C$^{2+}$ ion were convoluted with a 28~meV FWHM gaussian to simulate the experimental energy resolution and are shown separately in the four upper panels of figure~\ref{fig:C2overviewcomparison} for the different initial states relevant to the experiment. The lowest panel presents the measured data (solid line).

For quantitative comparison of the new calculations with the previous experimental data, figure~\ref{fig:C2overview} shows the  photoionization cross section measured at 28~meV energy spread and a theoretical model spectrum  obtained by assuming a metastable fraction $f_{\rm m}=0.5$ in the parent ion beam and statistical population of the (1s$^2$\,2s2p~$^3$P$^o_0$), (1s$^2$\,2s2p~$^3$P$^o_1$) and (1s$^2$\,2s2p~$^3$P$^o_2$) metastable states.  Evidently, these assumptions lead to a very satisfying agreement between theory and experiment. Given the total systematic uncertainty of $\pm 20$~\% of the experimental cross sections the metastable fraction $f_{\rm m}$ has an uncertainty as well. In the previous work on photoionization of C$^{2+}$ \cite{Mueller2002b} $f_{\rm m}$ was assumed to be $f_{\rm m}=0.4$ with 3/4 of that in the $^3$P$^o_0$ state and 1/8 each in the $^3$P$^o_1$ and $^3$P$^o_2$ states. These numbers must be attributed to the previous R-matrix calculations on an insufficient 1.36~meV energy mesh and to the experiment with its limited energy resolution.

\begin{figure}
\begin{center}
\includegraphics[width=\textwidth]{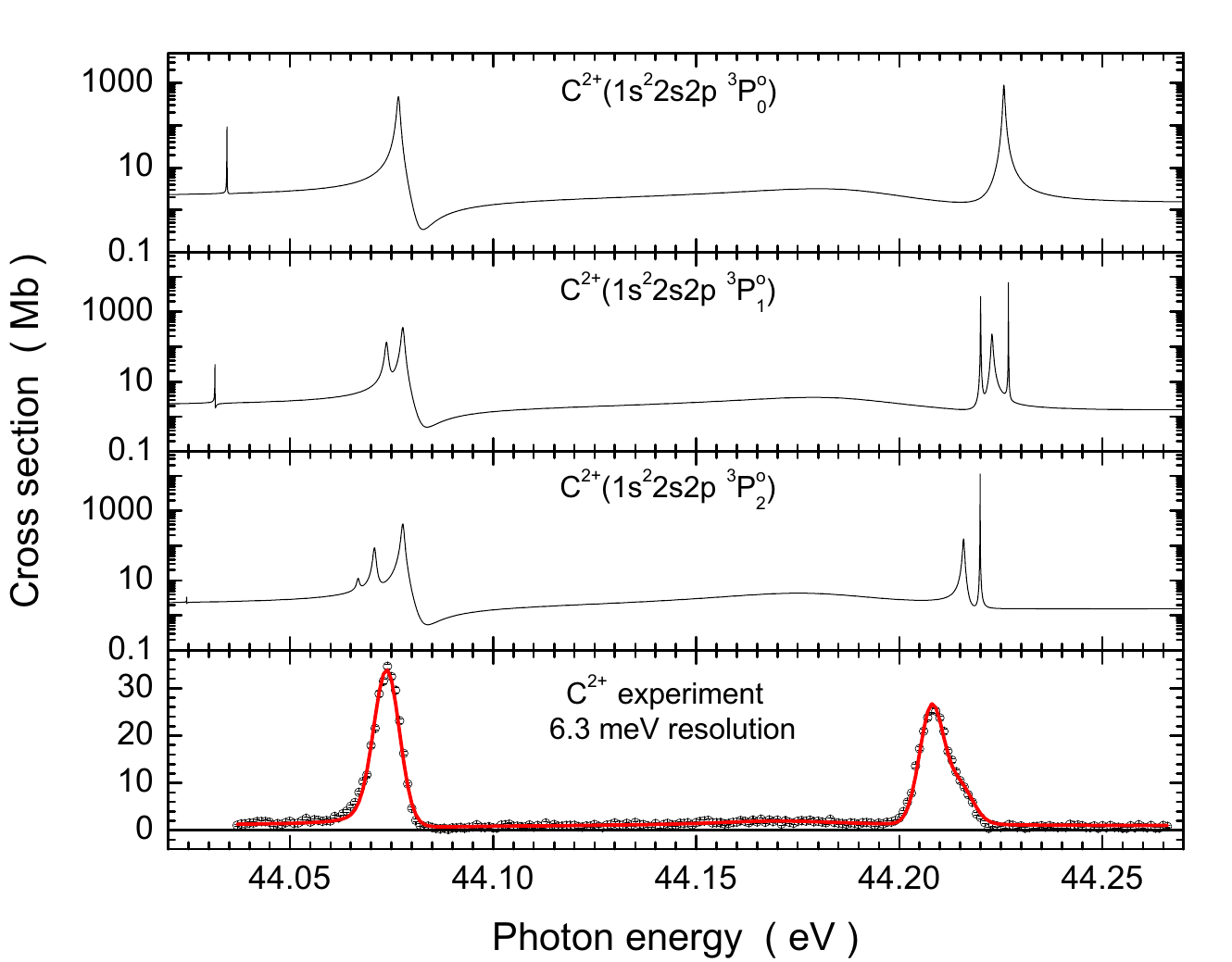}
\caption{\label{fig:C2first2} (Colour online) Results of the present R-matrix calculations in the energy range 44.00~to~44.27~eV for the $^3$P$^o$ initial states of the C$^{2+}$ ion with total angular momenta J=0,1, and 2 (first 3 panels). The bottom panel shows the data for a high-resolution photoionization scan in that energy range. The peak structure near 44.075~eV is associated with  excitations (2s2p $^3$P$^o$)$\to$(2p5p $^3$D) while the peak at 44.21~eV is related to excitations (2s2p $^3$P$^o$)$\to$(2p5p $^3$P). The solid (red) line is a fit to the data assuming the presence of only 4 contributing resonances in the experimental spectrum. This fit suggested an energy spread of 7~meV. Detailed comparison with the present theory results in an experimental energy resolution of 6.3~meV in this experiment. Note the logarithmic scale in the upper three panels.
}
\end{center}
\end{figure}

In the present study new measurements were conducted with the goal to push the energy resolution to its practical limits. For that purpose, the energy range 44.035-44.265~eV was scrutinized zooming in on the first two resonances seen in the overview spectrum in figure~\ref{fig:C2overview}. These peaks are associated with 2s$\to$np photoexcitations of the $^3$P$^o$ metastable levels of C$^{2+}$ leading to states with configuration (1s$^2$\,2p5p). Calculations of transition energies and absorption oscillator strengths $gf$ using the CATS code \cite{LANL,Cowan1981} showed that the first peak is made up of contributions to $^3$D$_{1,2,3}$ states while the second peak is associated with $^3$P$_{0,1,2}$ states. The detailed structure contained in the second resonance peak is already shown in figure~\ref{fig:C2theoryzoom}. For the energy range selected in the high-resolution experiment of this study the new calculations are shown in figure~\ref{fig:C2first2} for the $^3$P$^o$ initial states of the C$^{2+}$ ion with total angular momenta J=0,1, and 2 (first 3 panels). The bottom panel shows the data for a high-resolution photoionization scan in that energy range. The relative data of this scan were normalized to the previous absolute cross section measurement \cite{Mueller2002b}. The solid line is a fit to the data assuming the presence of only 4 contributing resonances in the experimental spectrum. This fit resulted in an overestimated energy spread of 7~meV. Detailed comparison with the present theory reveals an experimental energy resolution of 6.3~meV in this experiment.

\begin{figure}
\begin{center}
\includegraphics[width=\textwidth]{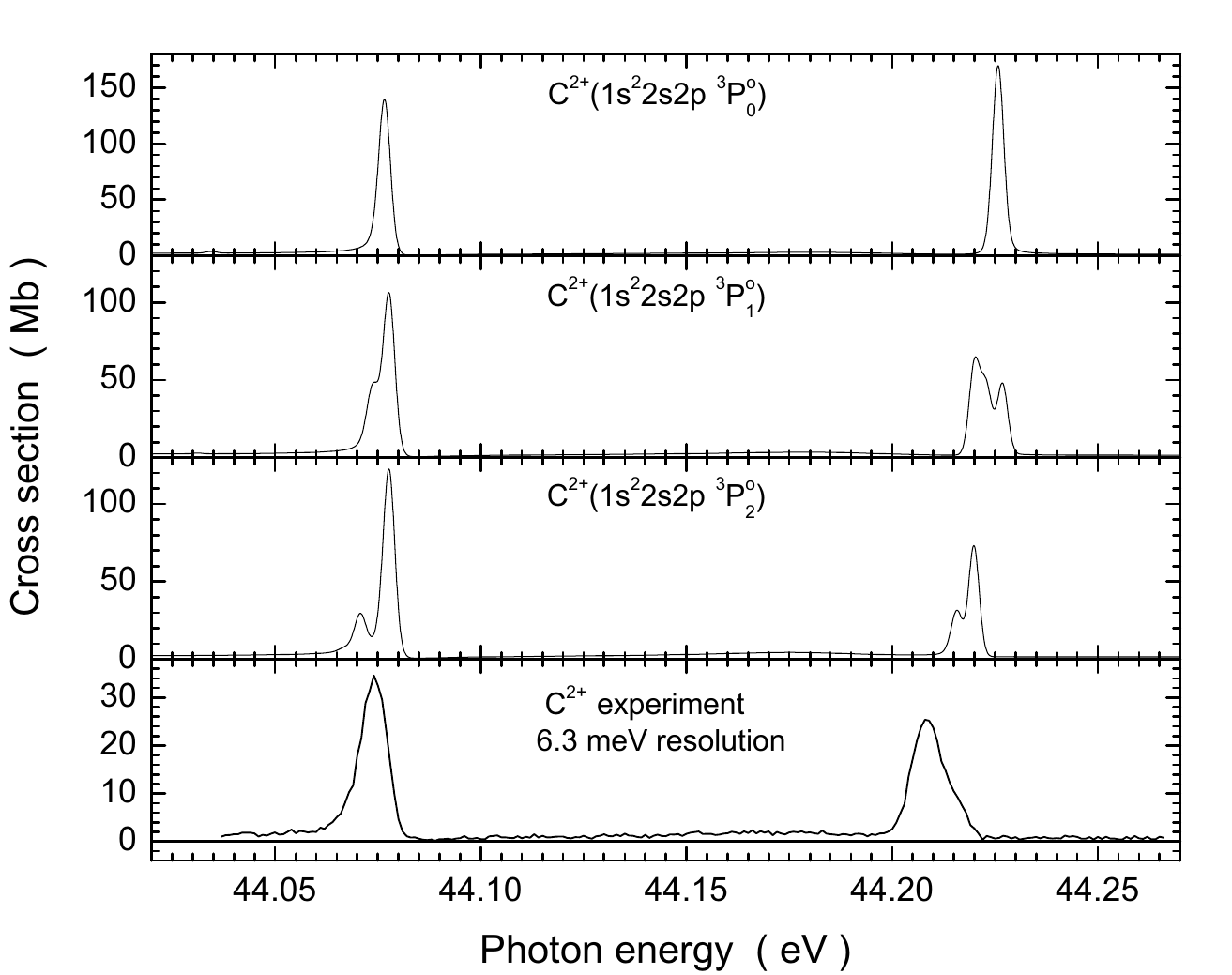}
\caption{\label{fig:C2first2convoluted} Simulated results of  experiments on photoionization of individual C$^{2+}$(2s2p~$^3$P$^o$) ions with 6.3~meV resolution in the energy range of figure~\ref{fig:C2first2}. The cross sections in all panels are on a linear scale. The bottom panel is the same as in figure~\ref{fig:C2first2}.
}
\end{center}
\end{figure}

In order to find out which of the peak structures in figure~\ref{fig:C2first2} contribute significantly to the experimental observation, the theoretical data were convoluted with a 6.3~meV FWHM gaussian. The results are shown in figure~\ref{fig:C2first2convoluted}. The cross sections in the sequence of the upper 3 panels are for pure parent ion beams each consisting of ions in (1s$^2$\,2s2p~$^3$P$^o_0$), (1s$^2$\,2s2p~$^3$P$^o_1$) or (1s$^2$\,2s2p~$^3$P$^o_2$) states, respectively. For the direct comparison with the experimental data, a weighted sum of the individual contributions from the three metastable components of the parent ion beam has to be calculated. From the comparison shown in figure~\ref{fig:C2overviewcomparison} one has to expect $f_{\rm m}=0.5$ for the total metastable fraction and statistical population, i.e., 1/9, 3/9, and 5/9 for the J=0, 1, and 2 metastable components, respectively.

\begin{figure}
\begin{center}
\includegraphics[width=\textwidth]{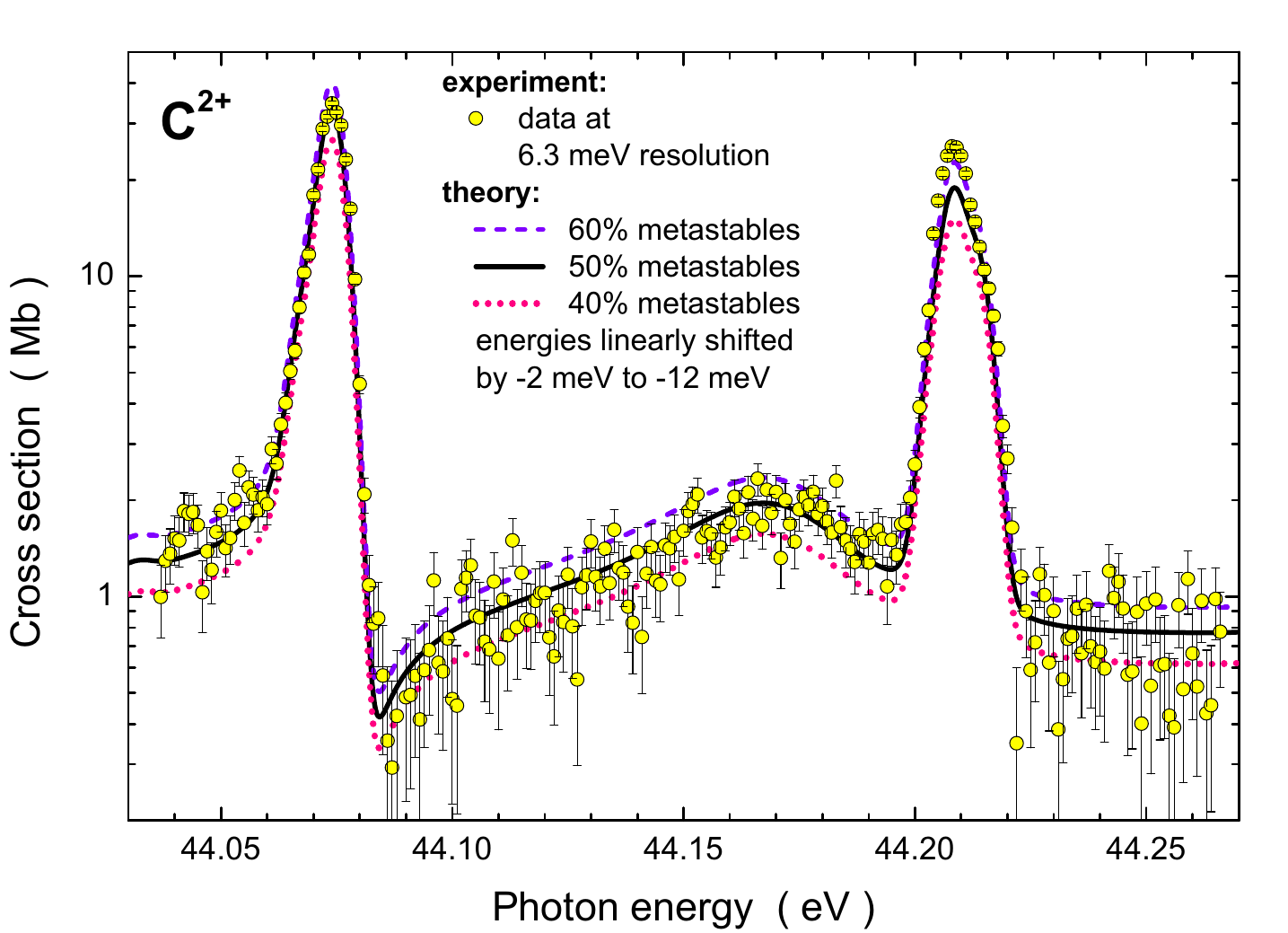}
\caption{\label{fig:C2first2comparisonlog}  (Colour online) Comparison of simulated and measured photoionization cross sections of C$^{2+}$ ions at an energy resolution of 6.3~meV in the energy range of Figs.~\ref{fig:C2first2} and~\ref{fig:C2first2convoluted}. The total fraction of metastable contributions was varied assuming $f_{\rm m} = 0.5$ (solid line), $f_{\rm m} = 0.6$ (dashed line) and $f_{\rm m} = 0.4$ (dotted line). The energy scale of the theoretical results was subjected to a linear shift function correcting the energy at the low-energy side of the theoretical spectrum by -2~meV and -12~meV at the high-energy side. Error bars displayed are statistical only.
}
\end{center}
\end{figure}

For the detailed comparison of the theoretically simulated cross section with the experiment figure~\ref{fig:C2first2comparisonlog} shows simulated and measured photoionization cross sections of C$^{2+}$ ions at an energy resolution of 6.3~meV in the energy range of figures~\ref{fig:C2first2} and~\ref{fig:C2first2convoluted}. The cross section scale is logarithmic putting emphasis on the small contributions from direct photoionization and on the interference patterns of resonant and direct photoionization channels. The total fraction of metastable contributions was varied assuming $f_{\rm m} = 0.5$ (solid line), $f_{\rm m} = 0.6$ (dashed line) and $f_{\rm m} = 0.4$ (dotted line). For easier comparison of theoretical and experimental data the energy scale of the theoretical results was shifted by a linear function of the photon energy $E_{\rm ph}$ in the rest frame of the ion correcting $E_{\rm ph}$ at the low-energy side of the theoretical spectrum by -2~meV and -12~meV at the high-energy side. Such shifts are outside the uncertainty of the present experiment and indicate the limitations in accuracy of the present structure calculations. With the small energy shifts introduced, the agreement between theory and experiment is truely striking. The error bars of the experiment leave some uncertainty, though, of the metastable fraction $f_{\rm m}$.

\begin{figure}
\begin{center}
\includegraphics[width=\textwidth]{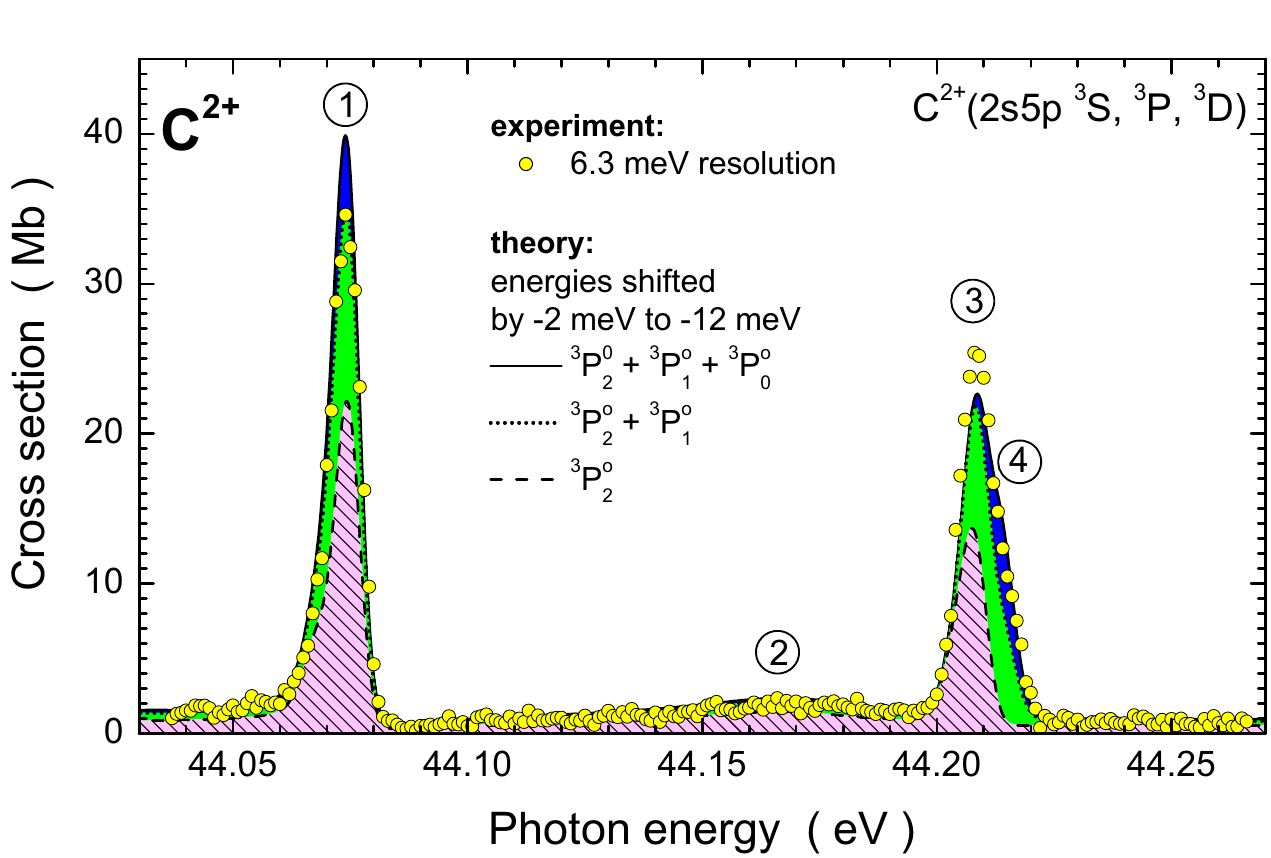}
\caption{\label{fig:C2first2comparisonlin} (Colour online) Illustration of the individual contributions of C$^{2+}$ ions in the (1s$^2$\,2s2p~$^3$P$^o_0$), (1s$^2$\,2s2p~$^3$P$^o_1$) and (1s$^2$\,2s2p~$^3$P$^o_2$) metastable states to the experimental high-resolution scan measurement shown also in the previous figures. The scan data are shown as open (yellow) shaded circles. The statistical error bars are within the size of the symbols. The total metastable fraction in the beam was set to $f_{\rm m}=0.6$ in the theoretical simulation. The 5/9 contribution of the $^3$P$^o_2$ fine structure state is shown by the dashed line (hatched area with red shading). On top of that is the 3/9 $^3$P$^o_1$ contribution represented by the difference between the dotted and the dashed line (the green shaded area). Further adding of the 5/9 $^3$P$^o_0$ contribution (the blue shaded area) results in the solid-line envelope of all theoretical contributions to the simulated cross section. The energy scale is the same as in Fig.~\ref{fig:C2first2comparisonlog}. The peak numbers indicate the 4 peaks assumed in the fit to the data as shown in figure~\ref{fig:C2first2}. Autoionizing (2p5p~$^3$S,$^3$P,$^3$D) states are formed in the present energy range.
}
\end{center}
\end{figure}

With the excellent agreement of theory and experiment demonstrated in figure~\ref{fig:C2first2comparisonlog}, a meaningful separation of the individual contributions of the (1s$^2$\,2s2p~$^3$P$^o_0$), (1s$^2$\,2s2p~$^3$P$^o_1$) or (1s$^2$\,2s2p~$^3$P$^o_2$) metastable states in the parent ion beam is possible. The separated individual contributions to the two first peak features of the photoionization spectrum are illustrated in figure~\ref{fig:C2first2comparisonlin}. This time, the cross section scale is linear. Apart from that the experimental data are those of figure~\ref{fig:C2first2comparisonlog}. For the comparison with theory a metastable fraction of $f_{\rm m}=0.6$ was assumed in the theoretical simulation of the experimental data, mainly because it gives the best agreement for the heights of the two peaks. One has to keep in mind here, that the fraction $f_{\rm m}$ is uncertain and its accurate determination would require a substantially increased effort. The experimental error bars provided in figure~\ref{fig:C2first2comparisonlin} are statistical only. The total systematic uncertainty is $\pm$~20\% as mentioned above.

\begin{figure}
\begin{center}
\includegraphics[width=0.9\textwidth]{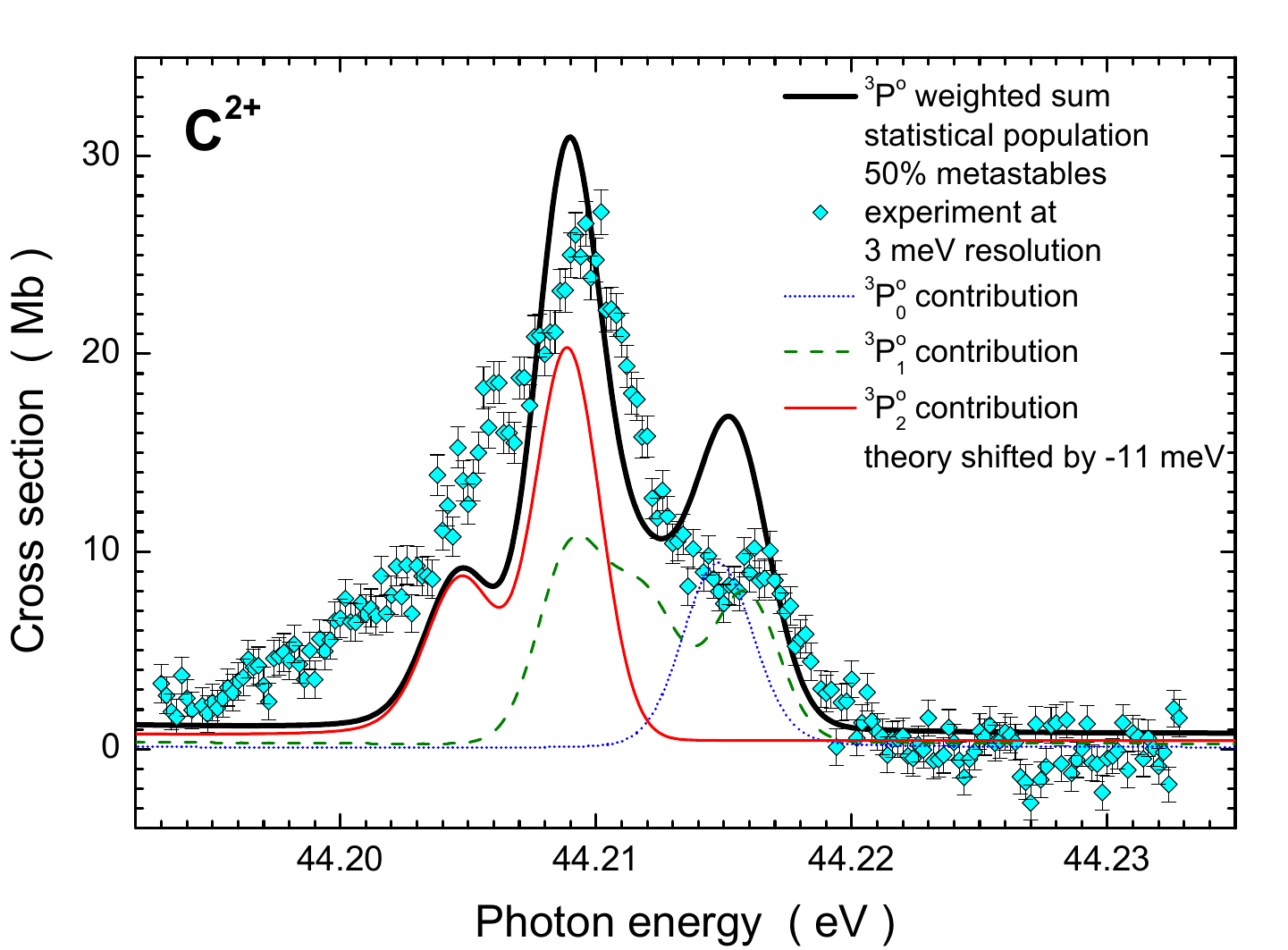}
\caption{\label{fig:C2second3meV} (Colour online) Comparison of theory and a scan measurement at 3~meV resolution  of the second peak from the last several figures. The experimental data are represented by open (cyan) shaded diamonds with statistical error bars. For the theoretical simulation of the experiment a total metastable fraction $f_{\rm m}=0.5$ was assumed. The theoretical data were shifted in energy by -11~meV. The contributions of the individual fine structure states were calculated assuming statistical weights. They are represented by the dotted line (the $^3$P$^o_0$ contribution), the dashed (green) line (the $^3$P$^o_1$ contribution), and the solid (red) line line (the $^3$P$^o_2$ contribution). The sum of these contributions is the fat solid (black) line representing the theoretical prediction for the measured cross section.
}
\end{center}
\end{figure}

Although the resolving power of the new experimental scan measurement is already as high as 7,000 it is not sufficient to really resolve individual contributions from the different $^3$P$^o$ fine structure states. Therefore an effort was made to further reduce the experimental energy spread. Figure~\ref{fig:C2second3meV} shows a comparison of theory and a scan measurement at 3~meV resolution  of the second peak from the last several figures. Now the resolving power is 14,700. For the theoretical simulation of the experiment a total metastable fraction $f_{\rm m}=0.5$ was assumed in this case. The theoretical data were shifted in energy by -11~meV. The contributions of the individual fine structure states were calculated assuming statistical weights. In spite of the energy shift of the theoretical data, the simulated cross section and the measurement do not fully line up with one another. Beside the overall shift there are additional slight deviations of the energies of resonances arising from individual $^3$P$^o$ fine structure states. In spite of the remaining small discrepancies it is seen that the experiment at this level of resolving power starts to separate the individual $^3$P$^o$ fine structure components of the parent ion beam. Although the agreement between theory and experiment is not perfect at this level of experimental precision, the structured shape of the investigated peak feature is very well reproduced by the present theory.

For gaining further insight and for supporting the present analysis, calculations of excitation energies and oscillator strengths relevant to the energy region of interest were carried out using the online version of the CATS code \cite{LANL,Cowan1981}. From the calculated data it is possible to get an impression of the individual contributions of specified excitation channels to the observed photoionization cross sections.
The photoionization resonance strengths for excitations from initial states $i$ to final states $f$
\begin{equation}\label{eq:sigmabar}
\overline{\sigma}_{i,f} = \int_{-\infty}^{\infty}  \sigma_{i,f} ({\rm E})d{\rm E}
\end{equation}
with the photoionization cross section $\sigma_{i,f} ({\rm E})$
can be calculated as the product of the oscillator strength for absorption $gf$ and the branching ratio for autoionization, i.\,e.,
\begin{equation}\label{eq:sigmafromgf}
  \overline{\sigma}_{i,f} = 4\pi^2 \alpha\, a_0^2 \,{\cal R}\, \frac{gf}{g_i} \omega_a
\end{equation}
with the fine structure constant $\alpha$, the Bohr radius $a_0$, the
Rydberg constant ${\cal R}$ and the statistical weight $g_i$ of the
initial state ($4\pi^2 \alpha\, a_0^2 \,{\cal R} \approx
109.76$~Mb~eV). The branching ratio $\omega_a$ can be assumed to be unity. Not regarding the effects of finite natural line widths and neglecting the influence of interference between resonant and direct photoionization, the oscillator strengths found for individual transitions are directly related to the individual contributions of these transitions to the experimental cross sections. To be more precise, the oscillator strengths allow one to obtain resonance strengths by employing equation~\ref{eq:sigmafromgf} and by using the appropriate weight factors for the initial states.

%
%
%

\begin{table}
\caption{\label{tab:LANLmodelC2} Assignments of the resonance features in the photoionization of C$^{2+}$ ions displayed in figure~\ref{fig:C2first2comparisonlin} inferred from results of a calculation employing the CATS code \cite{LANL,Cowan1981}. Note that the 3 groups of transitions had to be shifted by different amounts of energy to match the experiment. The ratios of oscillator strengths $gf$ are a measure of the individual contributions (see equation~\ref{eq:sigmafromgf}) by the indicated transitions.
}
\begin{indented}
 \lineup
 \item[]\begin{tabular}{ccr@{\,}c@{\,}llcl}
\br
 peak    & energy & \multicolumn{3}{c}{transition}   & \multicolumn{1}{c}{energy} & \multicolumn{2}{c}{$gf$}\\
 no.      & (exp., eV) & \multicolumn{3}{c}{(2s2p $\to$ 2p5p)} & \multicolumn{1}{c}{(CATS, eV)} & \multicolumn{2}{c}{(CATS)}\\
 \ns
 \mr
 1 & 44.074   &   $^3$P$^o_2$    &$\to$   &     $^3$D$_2$   &   44.068$^a$	    &  0.00784\\
   &          &   $^3$P$^o_1$    &$\to$  	&     $^3$D$_1$   &   44.070$^a$       	 &  0.00798\\
   &          &   $^3$P$^o_1$    &$\to$  	&     $^3$D$_3$   &   44.074$^a$       	 &  0.04701\\
   &          &   $^3$P$^o_0$    &$\to$  	&     $^3$D$_1$   &   44.074$^a$       	 &  0.01140\\
   &          &   $^3$P$^o_1$    &$\to$  	&     $^3$D$_2$   &   44.074$^a$       	 &  0.02573\\

\br

2  &  44.170  &   $^3$P$^o_2$    &$\to$  	&     $^3$S$_1$   &   44.169$^b$       	 &  0.01042\\
   &          &   $^3$P$^o_1$    &$\to$  	&     $^3$S$_1$   &   44.176$^b$       	 &  0.00722\\
   &          &   $^3$P$^o_0$    &$\to$  	&     $^3$S$_1$   &   44.180$^b$       	 &  0.00251\\

\br

3 &  44.208  &   $^3$P$^o_2$    &$\to$  	&     $^3$P$_1$   &   44.206$^c$       	 &  0.00926\\
  &          &   $^3$P$^o_1$    &$\to$  	&     $^3$P$_0$   &   44.209$^c$       	 &  0.00674\\
  &          &   $^3$P$^o_2$    &$\to$  	&     $^3$P$_2$   &   44.209$^c$       	 &  0.02583\\

4 &  44.215  &   $^3$P$^o_1$    &$\to$  	&     $^3$P$_1$   &   44.213$^c$     	 &  0.00478\\
  &          &   $^3$P$^o_0$    &$\to$  	&     $^3$P$_1$   &   44.216$^c$       	 &  0.00619\\
  &          &   $^3$P$^o_1$    &$\to$  	&     $^3$P$_2$   &   44.216$^c$       	 &  0.00786\\

\br
\end{tabular}
~\\
$^{a}$shifted by -0.277~eV\\
$^{b}$shifted by -0.240~eV\\
$^{c}$shifted by -0.322~eV\\

\end{indented}
\end{table}

For the initial (2s2p) configuration there are the 3 metastable $^3$P$^o$ fine structure states with total angular momentum quantum numbers J=0,1,2 and a short lived $^1$P component which is not present in the parent ion beam. For the (2p5p) final configuration there are 10 different states. Oscillator strengths $gf$ for the 14 dominant specified transitions from the initial $^3$P$^o$ states to the (2p5p) configuration were calculated. One obtains three groups of resonances clearly separated in energy. Comparison with the experiment shows that the (online-version) CATS calculations \cite{LANL,Cowan1981} predict resonance energies that are roughly 0.3~eV too high. The energy shifts are even different for the three different groups of resonances. The results are illustrated in table~\ref{tab:LANLmodelC2} where the experimental resonance energies of the 4-peaks fit (see figure~\ref{fig:C2first2}) to the experimental data taken at 6.3~meV resolution are compared with the shifted energies of the associated transitions. The oscillator strengths $gf$ are provided for the 14 dominant photoexcitation  channels. The agreement of these calculations with the measured cross sections is remarkably good. This provides confidence in the designation of the peaks seen in figure~\ref{fig:C2first2comparisonlin}. Apparently, the two dominant peak features are associated with $^3$D excited states for the low-energy group and $^3$P states for the high-energy group of resonances. In between there is oscillator strength from excitations resulting in the population of the  $^3$S$_1$ state. The broad feature seen in the experiment between the $^3$D  and $^3$P peaks (see figure~\ref{fig:C2first2comparisonlog}) does contain some considerable resonance strength (proportional to $gf$, see equation~\ref{eq:sigmafromgf}) which is distributed over a relatively wide energy range. Apparently the natural width of the  $^3$S$_1$ state is much larger than the widths of the excited $^3$D  and $^3$P states.

\subsection{Photoionization of N\,$^{3+}$}\label{sec:N3}

\begin{figure}[hb]
\begin{center}
\includegraphics[width=\textwidth]{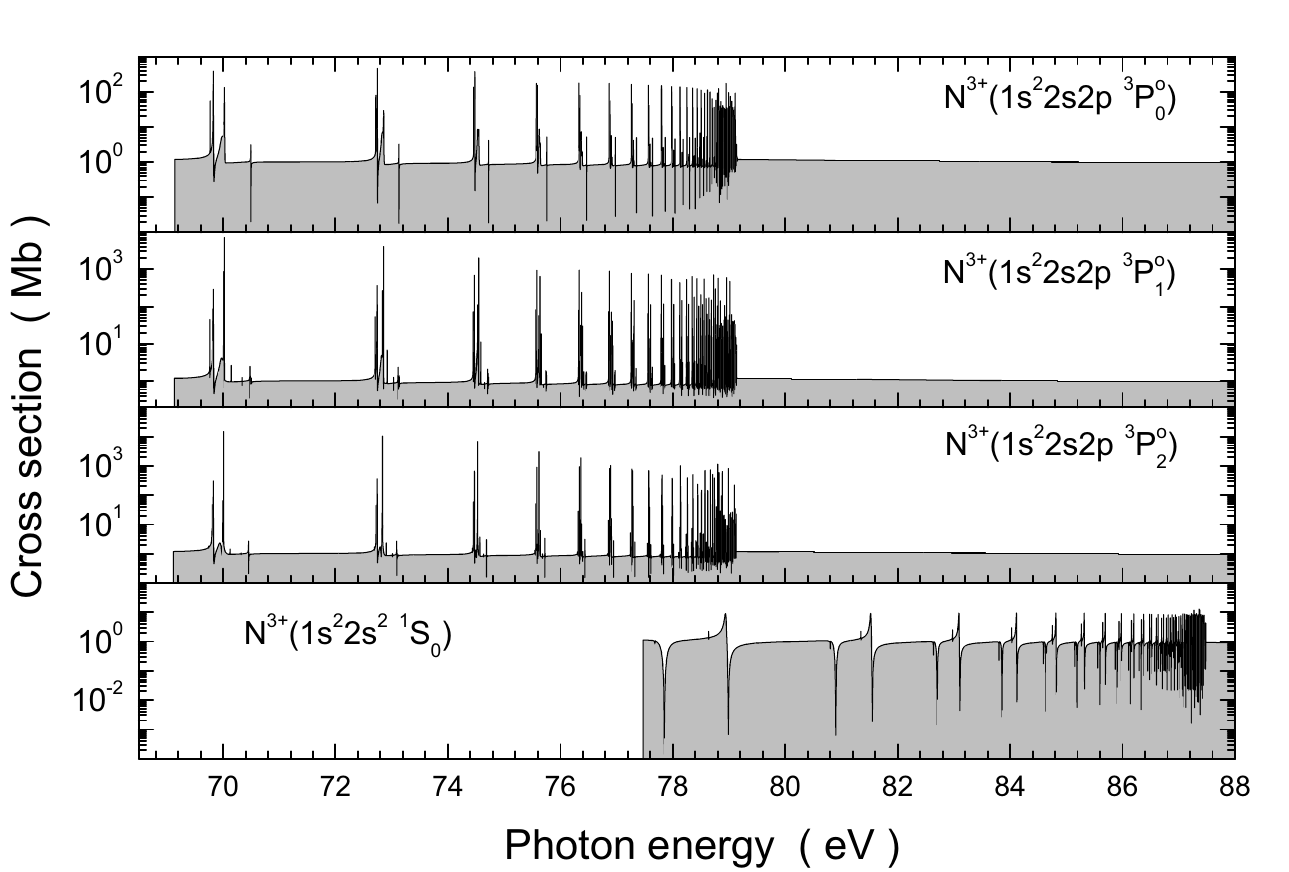}
\caption{\label{fig:N3theoryoverview} Results of the present Breit-Pauli R-matrix calculations for photoionization of N$^{3+}$ ions in the 4 different initial states: (1s$^2$\,2s2p~$^3$P$^o_0$) (top panel), (1s$^2$\,2s2p~$^3$P$^o_1$) (second panel from top), (1s$^2$\,2s2p~$^3$P$^o_2$) (third panel from top) and  (1s$^2$\,2s$^2$~$^1$S$_0$)(bottom panel). Note the logarithmic scale in each panel.
}
\end{center}
\end{figure}

\begin{figure}
\begin{center}
\includegraphics[width=\textwidth]{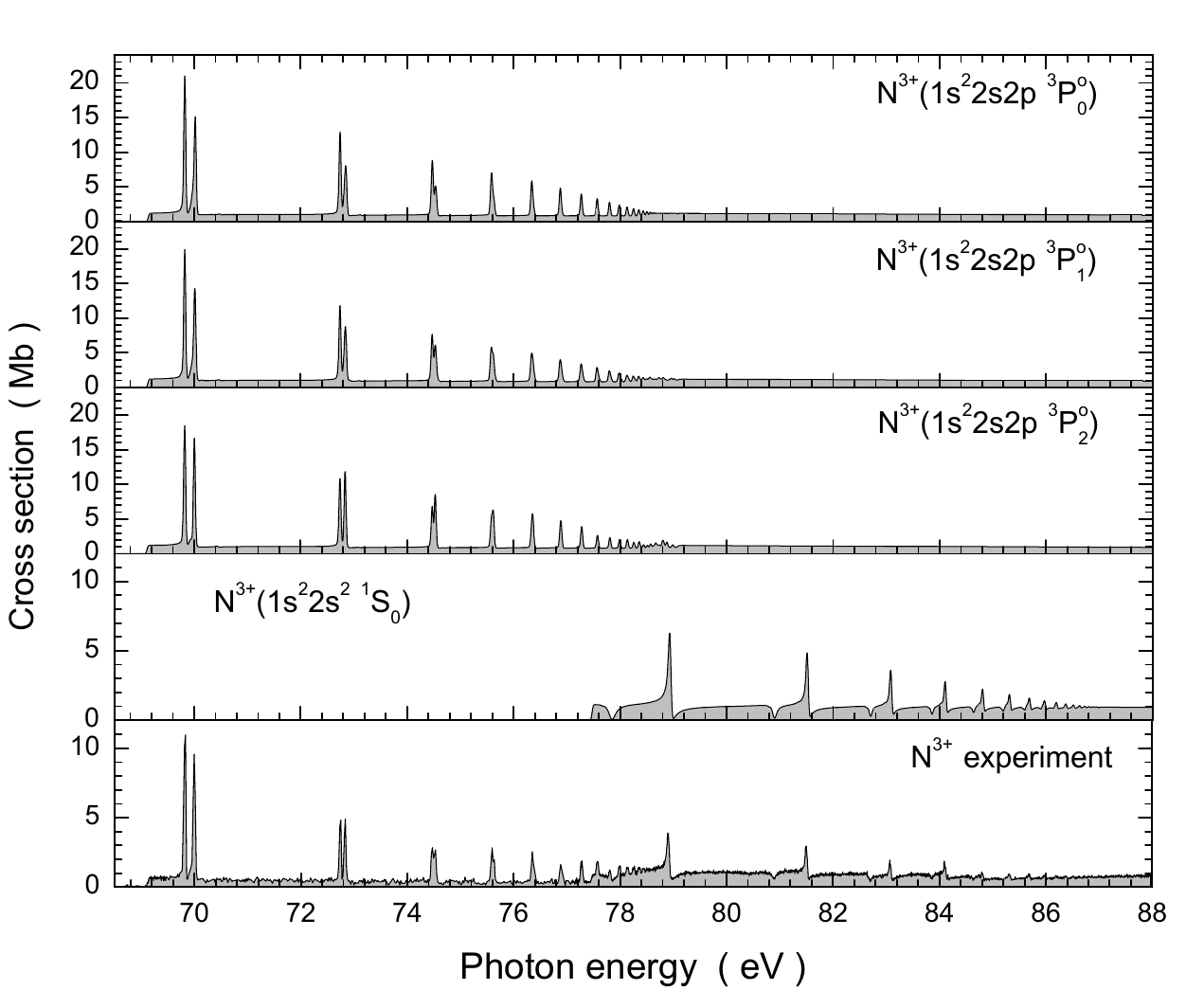}
\caption{\label{fig:N3theoryoverviewconvoluted} The theoretical data from figure~\ref{fig:N3theoryoverview} convoluted with a 40~meV FWHM gaussian and the experimental spectrum (lowest panel) measured  for photoionization of N$^{3+}$ ions. Note the linear scale in all five panels.
}
\end{center}
\end{figure}

Figure~\ref{fig:N3theoryoverview} shows the results of the present Breit-Pauli R-matrix calculations for photoionization of N$^{3+}$ ions in the 4 different (1s$^2$\,2s$^2$~$^1$S$_0$), (1s$^2$\,2s2p~$^3$P$^o_0$), (1s$^2$\,2s2p~$^3$P$^o_1$) and (1s$^2$\,2s2p~$^3$P$^o_2$) initial states. As in  the case of the C$^{2+}$ ion, all these initial states contribute to the experimental cross section obtained within the present new investigation on photoionization of Be-like ions. In a first measurement an overview photoionization spectrum was recorded for N$^{3+}$ ions covering an energy range between about 69~eV and 88~eV.  The energy resolution of this experiment is 40~meV. For comparison of the present calculations with the experiment the theoretical data were convoluted with a 40~meV FWHM gaussian to simulate the experimental energy resolution. The individual cross sections for the initial states relevant to the experiment are shown separately in the four upper panels of figure~\ref{fig:N3theoryoverviewconvoluted}. The bottom panel presents the measured data as points joined by a solid line. The measured relative cross section function was normalized to the absolute data of Bizau \etal~\cite{Bizau2005a}.

\begin{figure}
\begin{center}
\includegraphics[width=\textwidth]{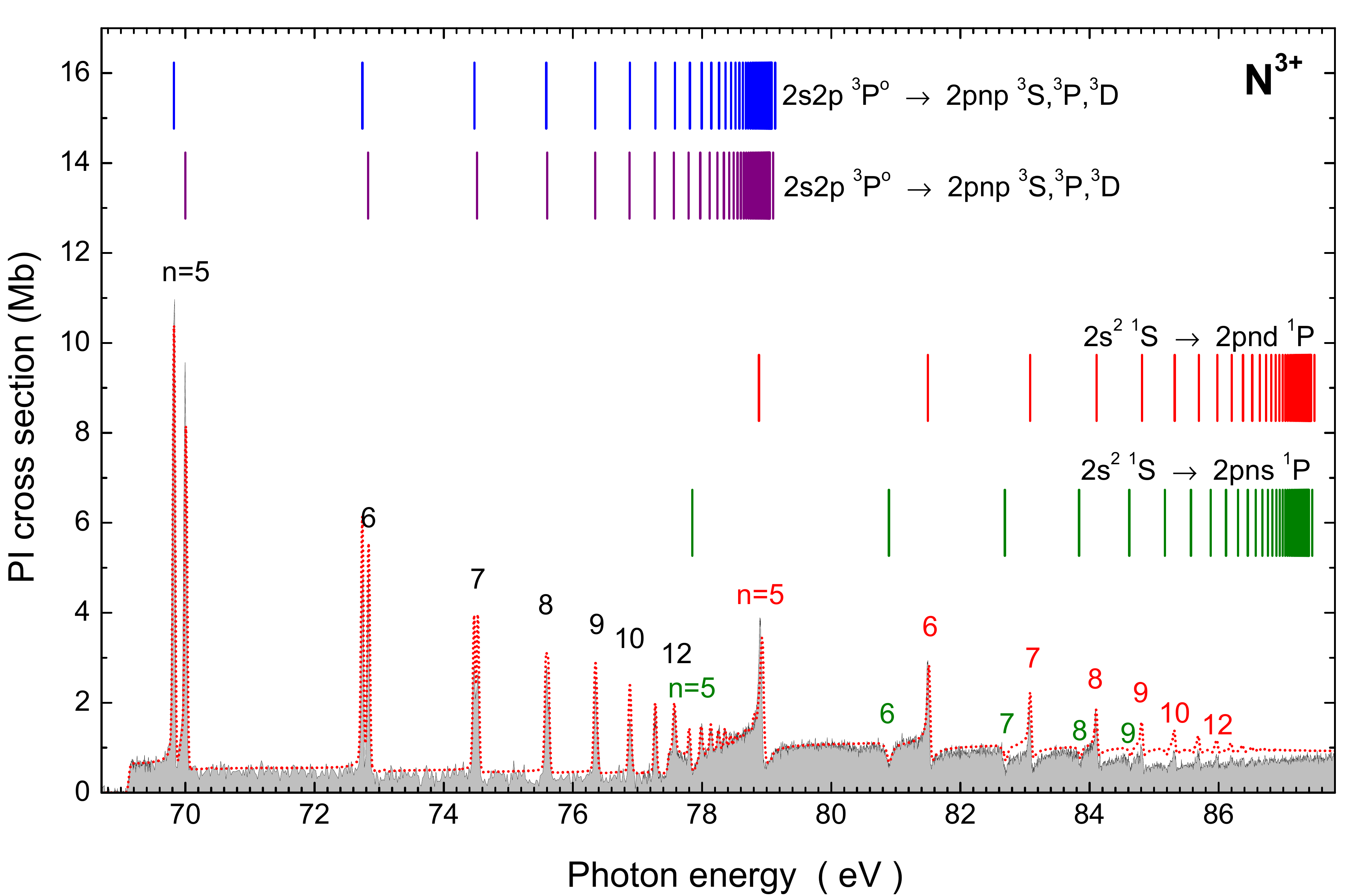}
\caption{\label{fig:N3overviewcomparison} (Colour online) Comparison of the experimental overview of N$^{3+}$ photoionization with a model spectrum obtained from the present Breit-Pauli R-matrix calculations. The theoretical spectrum was obtained by assuming a metastable fraction $f_{\rm m}=0.5$ in the parent ion beam and statistical population of the (1s$^2$\,2s2p~$^3$P$^o_0$), (1s$^2$\,2s2p~$^3$P$^o_1$) and (1s$^2$\,2s2p~$^3$P$^o_2$) metastable states. The experimental cross section is displayed as a solid line with light gray shading. The theoretical model spectrum is represented by the dotted (red) curve. The series of vertical lines indicate the Rydberg sequences of different groups of resonances excited by the photons as the photon energy increases. The (coloured) numbers on the resonances mark the principal quantum numbers of the associated Rydberg series.
}
\end{center}
\end{figure}

\begin{figure}
\begin{center}
\includegraphics[width=0.9\textwidth]{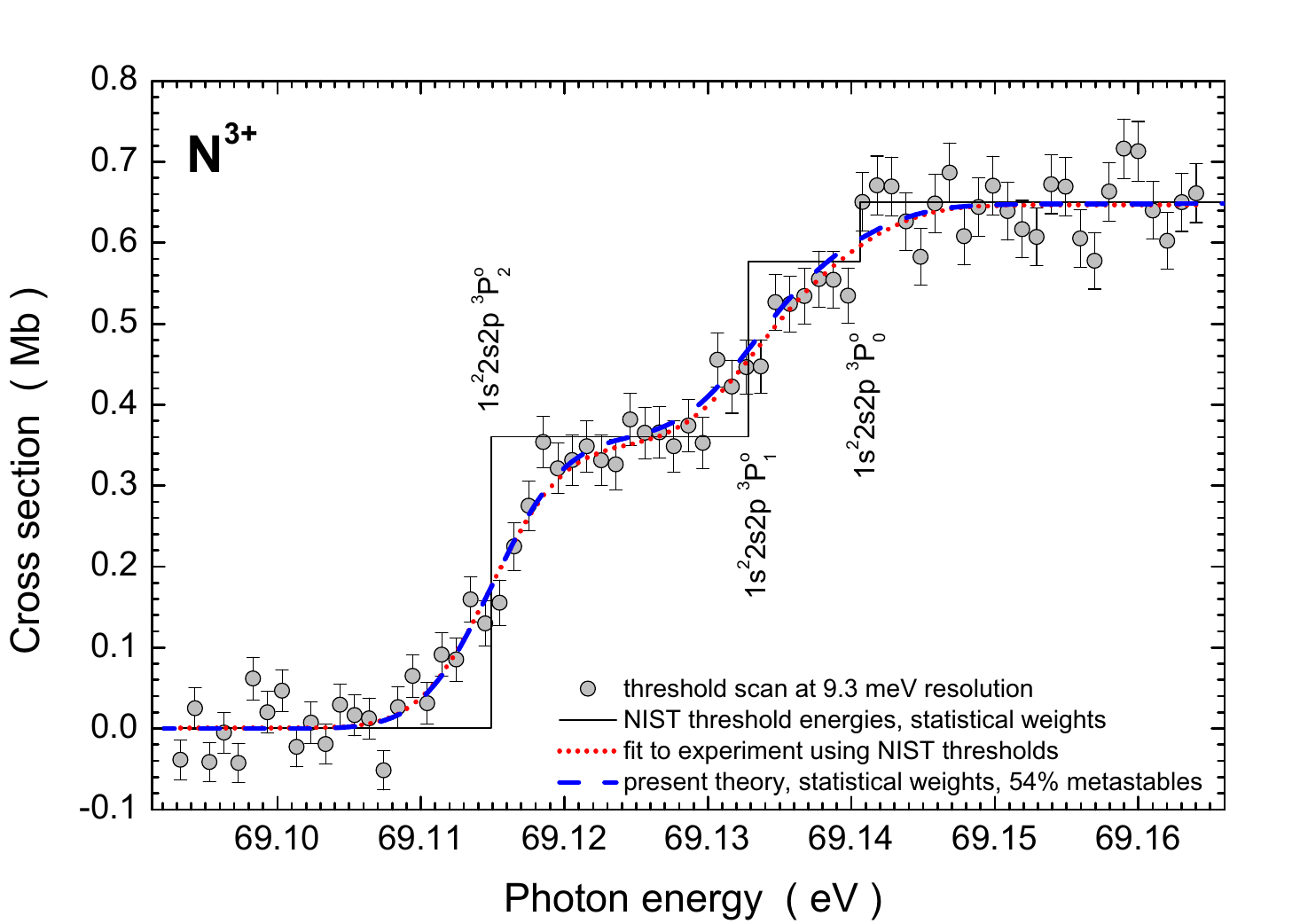}
\caption{\label{fig:N3threshold} (Colour online) Threshold scan for photoionization of N$^{3+}$ ions. The cross sections are normalized to the absolute measurements of Bizau \etal~\cite{Bizau2005a} and are represented by open circles with gray shading. Only statistical error bars are shown. The step function (solid black line) was constructed by using the ionization thresholds \cite{NIST2004} of the individual $^3$P$^o$ fine structure states and distributing the height of the threshold step (0.65~Mb) over contributions of the fine structure states assuming statistical population. The dotted (red) line is a fit to the experiment using the known ionization thresholds from reference~\cite{NIST2004} and fitting the individual step heights as well as the experimental energy spread to the measured spectrum.  The dashed (blue) line is a convolution of the theoretical result with a 9.3~meV FWHM gaussian. For the theoretical simulation a metastable fraction $f_{\rm m}=0.54$ and statistical population of the fine structure states was assumed.
}
\end{center}
\end{figure}

\begin{figure}
\begin{center}
\includegraphics[width=\textwidth]{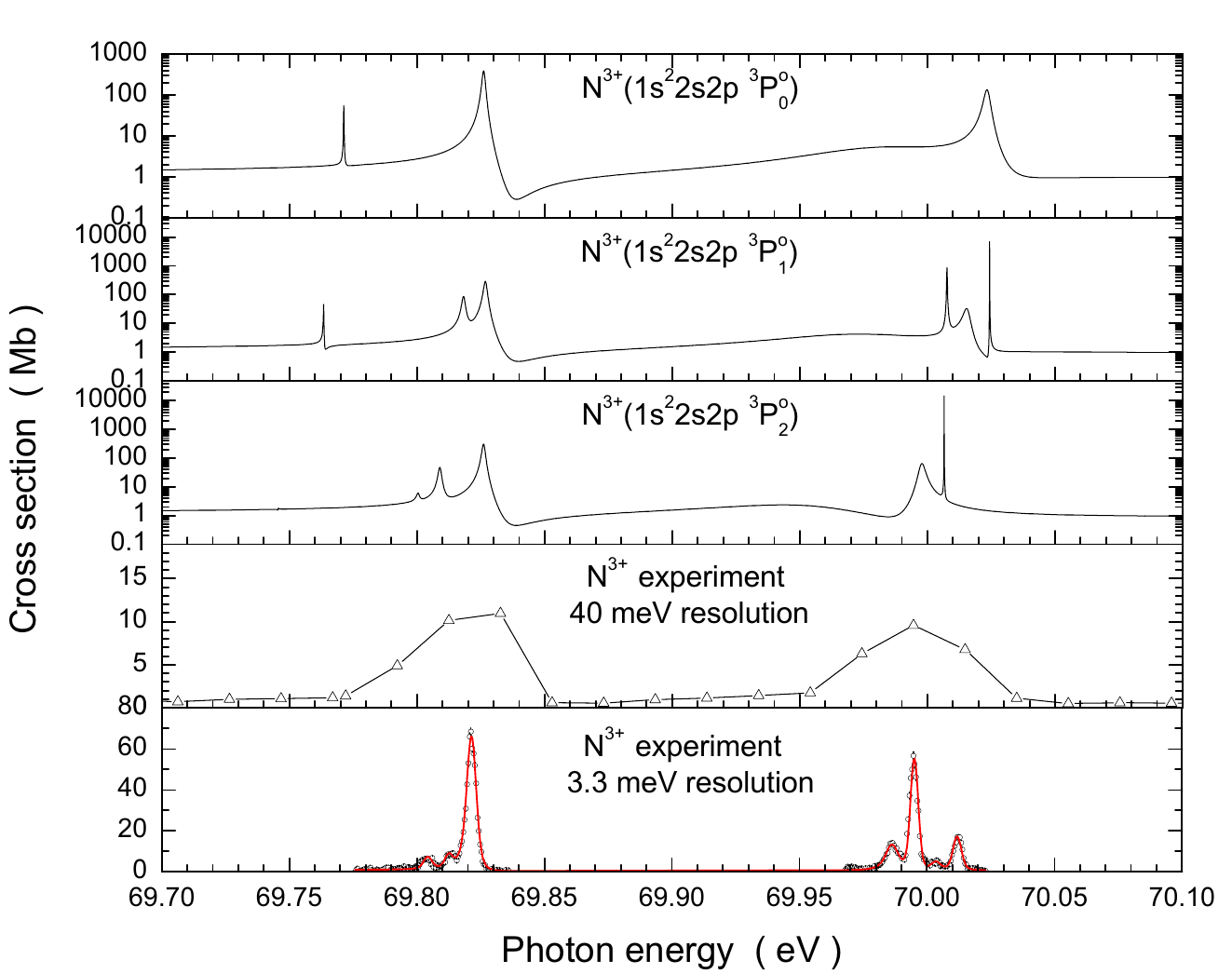}
\caption{\label{fig:N3first2} (Colour online) Results of the present R-matrix calculations in the energy range 69.7~to~70.1~eV for the $^3$P$^o$ initial states of the N$^{3+}$ ion with total angular momenta J=0, J=1, and J=2 (first 3 panels). The fourth panel displays the experimental cross section from the overview spectrum shown in figure~\ref{fig:N3overviewcomparison} at 40~meV resolution. The bottom panel presents a high-resolution photoionization scan in that energy range. The solid (red) line is a fit to the data assuming the presence of 7 contributing resonances in the experimental spectrum. This fit suggested an energy spread of about 4~meV. Detailed comparison with the present theory reveals an experimental energy resolution of 3.3~meV in this experiment. Note the logarithmic scale in the upper three panels.
}
\end{center}
\end{figure}

\begin{figure}
\begin{center}
\includegraphics[width=\textwidth]{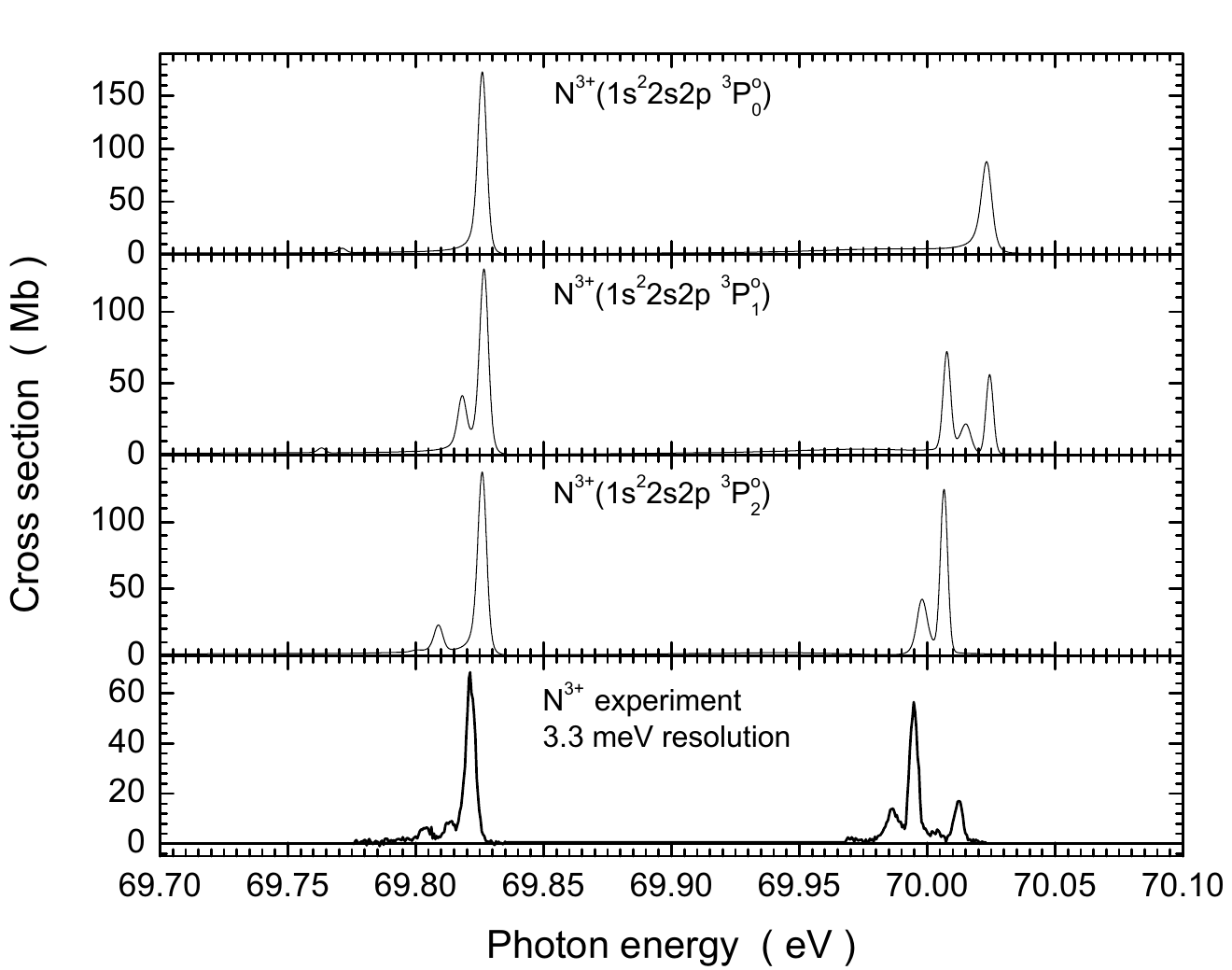}
\caption{\label{fig:N3first2convoluted} Simulated results of  experiments on photoionization of individual N$^{3+}$(2s2p~$^3$P$^o$) ions with 3.3~meV resolution in the energy range of figure~\ref{fig:N3first2}. The cross sections in all four panels are on a linear scale. The bottom panel shows the same experimental data as in figure~\ref{fig:N3first2}, this time as a solid line.
}
\end{center}
\end{figure}

\begin{figure}
\begin{center}
\includegraphics[width=\textwidth]{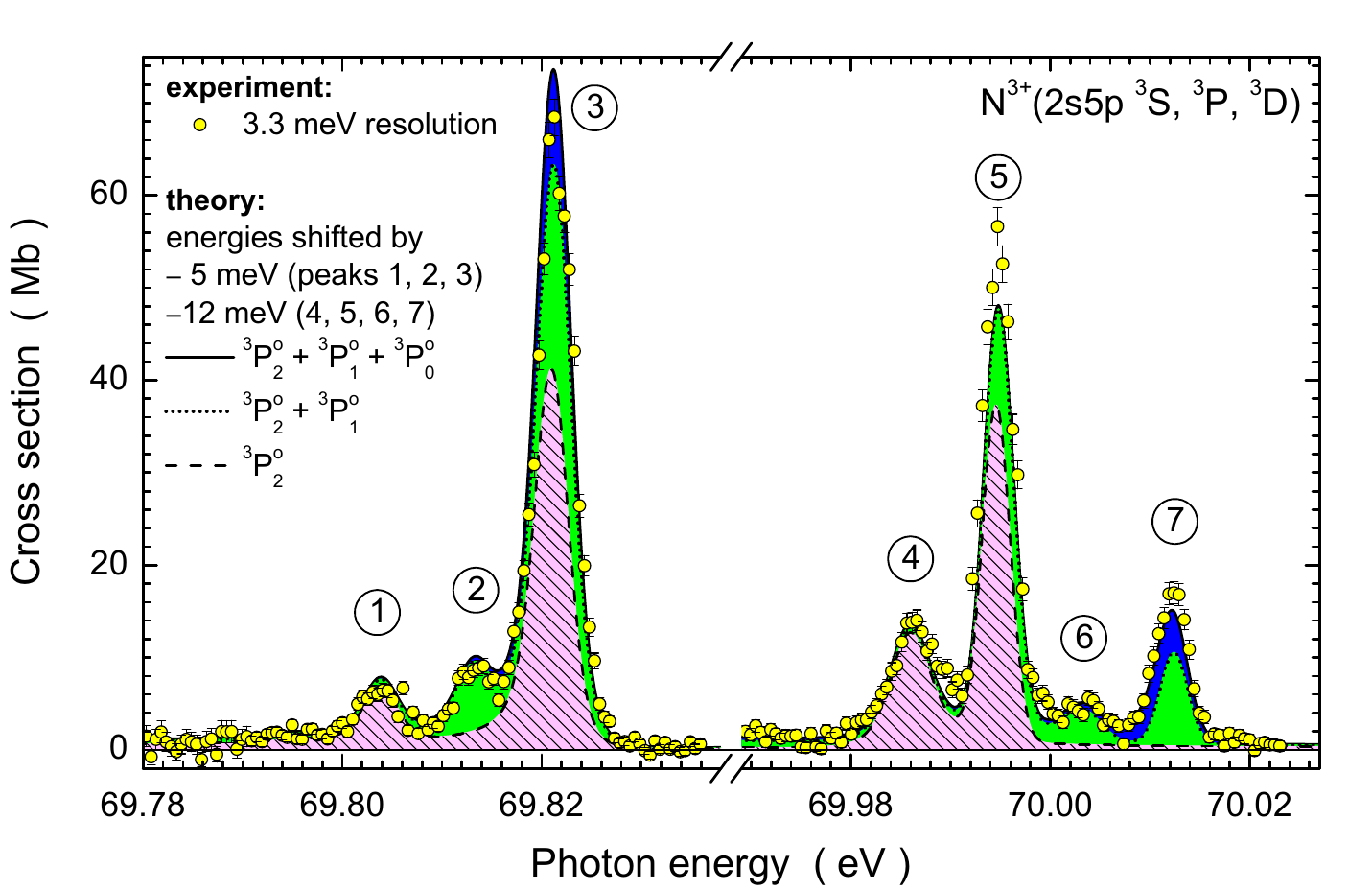}
\caption{\label{fig:N3first2comparison} (Colour online) Illustration of the individual contributions of ions in the (1s$^2$\,2s2p~$^3$P$^o_0$), (1s$^2$\,2s2p~$^3$P$^o_1$) and (1s$^2$\,2s2p~$^3$P$^o_2$) metastable N$^{3+}$ states to the experimental high-resolution scan measurement shown already in the previous figures. The scan data are displayed as open (yellow) shaded circles with statistical error bars. The total metastable fraction in the beam was set to $f_{\rm m}=0.54$ in the theoretical simulation. The 5/9 contribution of the $^3$P$^o_2$ fine structure state is shown by the dashed line (hatched area with red shading). On top of that is the 3/9 $^3$P$^o_1$ contribution represented by the difference between the dotted and the dashed line (the green shaded area). Further adding of the 1/3 $^3$P$^o_0$ contribution (the blue shaded area) results in the solid-line envelope of all theoretical contributions to the simulated cross section. The energy scale of the theoretical results was shifted by -5~meV for the whole lower energy group before the break in the energy axis and -12~meV for the whole higher energy group above the break. The encircled numbers mark the observed resonance peaks which are due to autoionizing (2p5p~$^3$S,$^3$P,$^3$D) states.
}
\end{center}
\end{figure}

Figure~\ref{fig:N3overviewcomparison} provides a comparison of the 40-meV resolution experimental overview of photoionization of N$^{3+}$ ions with a model spectrum obtained from the present Breit-Pauli R-matrix calculations. The theoretical spectrum was obtained by assuming a metastable fraction $f_{\rm m}=0.5$ in the parent ion beam and statistical population of the (1s$^2$\,2s2p~$^3$P$^o_0$), (1s$^2$\,2s2p~$^3$P$^o_1$) and (1s$^2$\,2s2p~$^3$P$^o_2$) metastable states. Theory and experiment are in very satisfactory agreement. At energies between approximately 84 and 86~eV, the measured cross sections are slightly below the theoretical data. This is assumed to be an artifact of the experimental data taking procedure. For obtaining the spectrum displayed in figure~\ref{fig:N3overviewcomparison} the total energy range of interest was divided up into about 20 individual overlapping scan ranges for each of which constant beam overlap form factors and constant photon energy spreads were assumed. Entrance and exit slits of the monochromator were only set for a given resolution at the center energy of each scan. They were not individually adjusted to each individual energy step. The total experimental spectrum was assembled by adjusting the relative measurements of adjacent scan ranges to the mutual overlap regions and thus combining the pieces step by step. This procedure may result in small excursions of the cross section over extended energy ranges.

The final states that can be reached by photo single excitation of the (2s2p $^3$P$^o$) metastable states via electric dipole transitions are members of the (2pnp $^3$S, $^3$P, $^3$D) Rydberg series. The associated series limits are the (2p $^2$P$_{1/2,\, 3/2}$) states of the Li-like product ion. For each of these two states Rydberg series with their resonance energies can be constructed using quantum defects. From a fit to our experimental data we find that the dominating series have quantum defects 0.110 and 0.165. The resonance energies can be described approximately by E$_n^{(1)}$ = 79.0998~eV - 4$^2 \times $ 13.6057~eV/(n-0.11)$^2$ and E$_n^{(2)}$ = 79.1318~eV - 4$^2 \times $ 13.6057~eV/(n-0.165)$^2$, respectively. The associated energies are shown by the vertical (coloured) bars above the data, the principal quantum numbers $n$ are provided to characterize the observed resonance features associated with the initial metastable $^3$P$^o$ states.

From the $^1$S ground state autoionizing resonant states can only be reached by two-electron excitations in this photon energy range. The final states that can be reached from the (2s$^2$ $^1$S) ground state are (2pns $^1$P) and (2pnd $^1$P) states. Again, two series of Rydberg states are involved converging towards the series limits (2p $^2$P$_{1/2,\, 3/2}$), shifted approximately by 8.35~eV (the average excitation energy of the $^3$P$^o$ metastable states)
with respect to the identical (2p $^2$P$_{1/2,\, 3/2}$) states when reached from the $^3$P$^o$ metastable initial states.
The two associated resonance energies of the dominating series  can be described approximately by E$_n^{(1)}$ = 87.4495~eV - 4$^2 \times $ 13.6057~eV/(n-0.239)$^2$ for the $^2$P$_{1/2}$ series limit  and E$_n^{(2)}$ = 87.4816~eV - 4$^2 \times $ 13.6057~eV/(n+0.03)$^2$ for the $^2$P$_{3/2}$ series limit. The associated energies are shown by vertical (coloured) bars above the data, and the principal quantum numbers $n$ are provided to characterize the observed resonance features associated with the initial  $^1$S ground state.
The (2pns $^1$P) states form window resonances in the photoionization spectrum. The (2pnd $^1$P) states form the strongest resonances in the photoionization spectrum. They interfere with direct photoionization of  the (2s$^2$ $^1$S) ground state, giving rise to strongly asymmetric Fano-Beutler peak features.

Figure~\ref{fig:N3threshold} shows a threshold scan for photoionization of N$^{3+}$ ions. The relative scan cross sections are normalized to the absolute measurements of Bizau \etal~\cite{Bizau2005a}. The simplest cross section model to represent the photoionization threshold is a step function with three steps, each associated with one of the three $^3$P$^o_{\rm J}$ metastable initial states.  The step function shown in the figure was constructed by using the ionization thresholds (77.4735~eV - 8.33288~eV / 8.34070~eV / and  8.35856~eV for J = 0, 1, and 2, respectively) \cite{NIST2004} of the individual $^3$P$^o$ fine structure states and distributing the height of the threshold step (0.65~Mb) over contributions of these fine structure states assuming statistical population.
A fit of a gaussian-convoluted step function with variable step heights to the experiment using the known ionization thresholds provides step heights consistent with a statistical population of the $^3$P$^o$ states. The fit also determined the energy spread of this threshold scan experiment to be 9.3~meV. For the theoretical simulation a metastable fraction $f_{\rm m}=0.54$ and statistical population of the fine structure states was assumed. The fit as well as  the theoretical result convoluted with a 9.3~meV FWHM gaussian give an excellent representation of the threshold cross section function. Note that the resolving power of the threshold scan measurement was more than 7,400. The good agreement between experiment and the model spectra strongly supports the assumption of statistical population of the metastable N$^{3+}$(2s2p $^3$P$^o$) states.

Figure~\ref{fig:N3first2} shows results of the present R-matrix calculations in the energy range 69.7~to~70.1~eV for the $^3$P$^o_{\rm J}$ initial states of the N$^{3+}$ ion with total angular momenta J=0, 1, and 2 ( see the first 3 panels). The theoretical data are to be compared with the experimental cross section from the overview spectrum displayed in figure~\ref{fig:N3overviewcomparison} at 40~meV resolution shown here in the fourth panel and the data from a high-resolution photoionization scan in that energy range displayed in the bottom panel. The solid (red) line is a fit to the data assuming the presence of 7 contributing resonances in the experimental spectrum. This fit suggested an energy spread of about 4~meV. Detailed comparison with the present theory reveals an experimental energy resolution of 3.3~meV in this experiment. Note the logarithmic scale in the upper three panels.

For the detailed direct comparison with experiment the theoretical contributions have to be convoluted with 3.3~meV FWHM gaussians simulating the energy spread in the measurement. The results of such convolution are provided in figure~\ref{fig:N3first2convoluted} and compared with the experiment (bottom panel). The next step is the modelling of the experiment by calculating the observed apparent cross section as a weighted sum of individual contributions from the different parent ion beam components. As in the analysis of the threshold energy scan (Figure~\ref{fig:N3threshold}), the metastable fraction $f_{\rm m}=0.54$ was assumed. The weights are 1/9 for the (1s$^2$\,2s2p~$^3$P$^o_0$) contribution, 3/9 for the (1s$^2$\,2s2p~$^3$P$^o_1$) contribution, and 5/9 for the (1s$^2$\,2s2p~$^3$P$^o_2$) as in all other cases shown in this paper. In figure~\ref{fig:N3first2comparison} the experimental data are compared to the theoretical model spectrum obtained by this procedure. In order to provide optimum conditions for detailed comparison,
the energy scale of the theoretical results was shifted by -5~meV for the whole lower energy group at around 69.8~eV and by -12~meV for the whole higher energy group at around 70~eV. The figure distinguishes between the contributions of individual initial metastable components. Seven distinct peak features can be seen. This is due to the resolving power of about 21,200 in this particular experiment. For all the other ions a correspondingly low energy spread could not be achieved. It is worthwhile, therefore, to take an especially detailed look at the features observed here. For supporting the present analysis, calculations of excitation energies and oscillator strengths relevant to the present region of interest were carried out using the CATS code \cite{LANL,Cowan1981}. As in the case of C$^{2+}$ the photoexcitation transitions from the initial metastable (2s2p $^3$P$^o$) levels to the excited states with a (2p5p) configuration were calculated. Oscillator strengths $gf$ for the 14 dominant transitions were obtained. Again three groups of resonances are clearly separated from one another.

%
%
%

\begin{table}
\caption{\label{tab:LANLmodelN3} Assignments of the resonance features in N$^{3+}$ ions displayed in figure~\ref{fig:N3first2comparison} inferred from
 calculations employing the CATS code \cite{LANL,Cowan1981}. The ratios of oscillator strengths $gf$ indicate the relative individual contributions of the specified transitions. Note the energy shifts of the theoretical data required to match the experiment.}
\begin{indented}
 \lineup
 \item[]\begin{tabular}{ccr@{\,}c@{\,}llcl}
\br
 peak    & energy & \multicolumn{3}{c}{transition}   & \multicolumn{1}{c}{energy} & \multicolumn{2}{c}{$gf$}\\
 no.      & (exp., eV) & \multicolumn{3}{c}{(2s2p $\to$ 2p5p)} & \multicolumn{1}{c}{(CATS, eV)} & \multicolumn{2}{c}{(CATS)}\\
 \ns
 \mr
 1 & 69.804	    & $^3$P$^o_2$     &$\to$  	& $^3$D$_2$  & 69.805$^a$	   & 0.00900   \\

2 &  69.813     & $^3$P$^o_1$     &$\to$  	& $^3$D$_1$  & 69.813$^a$      & 0.00896   \\

3 &  69.822     & $^3$P$^o_2$     &$\to$  	& $^3$D$_3$  & 69.821$^a$      & 0.05784   \\
  &             & $^3$P$^o_0$     &$\to$  	& $^3$D$_1$  & 69.821$^a$      & 0.01416   \\
  &             & $^3$P$^o_1$     &$\to$  	& $^3$D$_2$  & 69.821$^a$      & 0.03228   \\

\br

4 &  69.986     & $^3$P$^o_2$     &$\to$  	& $^3$P$_1$   & 69.984$^b$      &  0.01259  \\

5 &  69.995     & $^3$P$^o_2$     &$\to$  	& $^3$P$_2$   & 69.996$^b$      &  0.03238  \\
  &             & $^3$P$^o_1$     &$\to$  	& $^3$P$_0$   & 69.996$^b$     	&  0.00828  \\

6 &  70.004     & $^3$P$^o_1$     &$\to$  	& $^3$P$_1$   & 70.004$^b$      &  0.00541  \\

7 &  70.012     & $^3$P$^o_0$     &$\to$  	& $^3$P$_1$   & 70.012$^b$      &  0.00686  \\
  &             & $^3$P$^o_1$     &$\to$  	& $^3$P$_2$   & 70.012$^b$     	&  0.00896  \\

\br
\end{tabular}
~\\
$^{a}$shifted by -0.286~eV\\
$^{b}$shifted by -0.330~eV\\

\end{indented}
\end{table}

The relative strengths within each of the three resonance groups match the experiment extremely well, given the simplicity of the approach. In contrast to that, the relative strengths of the three different resonance groups with respect to one another do not agree equally well with the measured data. Also the resonance energies are above the experimental values by about 0.3~eV and discrepancies differ from one resonance group to the other. Since the experiment does not provide much evidence for the presence of (2p5p $^3$S$_1$) excited states and the possible resonance strengths of such contributions cannot be experimentally assessed, the $^3$S channels  are not included in table~\ref{tab:LANLmodelN3}.  Considering the extremely high resolution of the N$^{3+}$ experiment at 3.3~meV energy spread, the CATS calculations are in remarkably good accord with the experiment, providing high confidence in the designation of the peaks numbered in figure~\ref{fig:N3first2comparison}.

Simon \etal \cite{Simon2010} have observed the resonance group shown in figure~\ref{fig:N3first2comparison} at a resolving power of approximately 2,000. With an energy spread of 35~meV they could not resolve any of the fine details revealed by the present study with an energy spread as small as 3.3~meV. Their $^3$D and $^3$P peak assignments for the two resonance features observed agree with the present, however, the total angular momentum associated with the two peaks is not just J=1 as assumed by Simon \etal but results from various combinations of $^3$P$^o_{\rm J} \to ^3$D$_{\rm J'}$ and $^3$P$^o_{\rm J} \to ^3$P$_{\rm J''}$ photoexcitations with J=0,1,2, J'=1,2,3 and J''=0,1,2, respectively, as the present investigation demonstrates.

\subsection{Photoionization of O\,$^{4+}$}\label{sec:O4}

\begin{figure}
\begin{center}
\includegraphics[width=\textwidth]{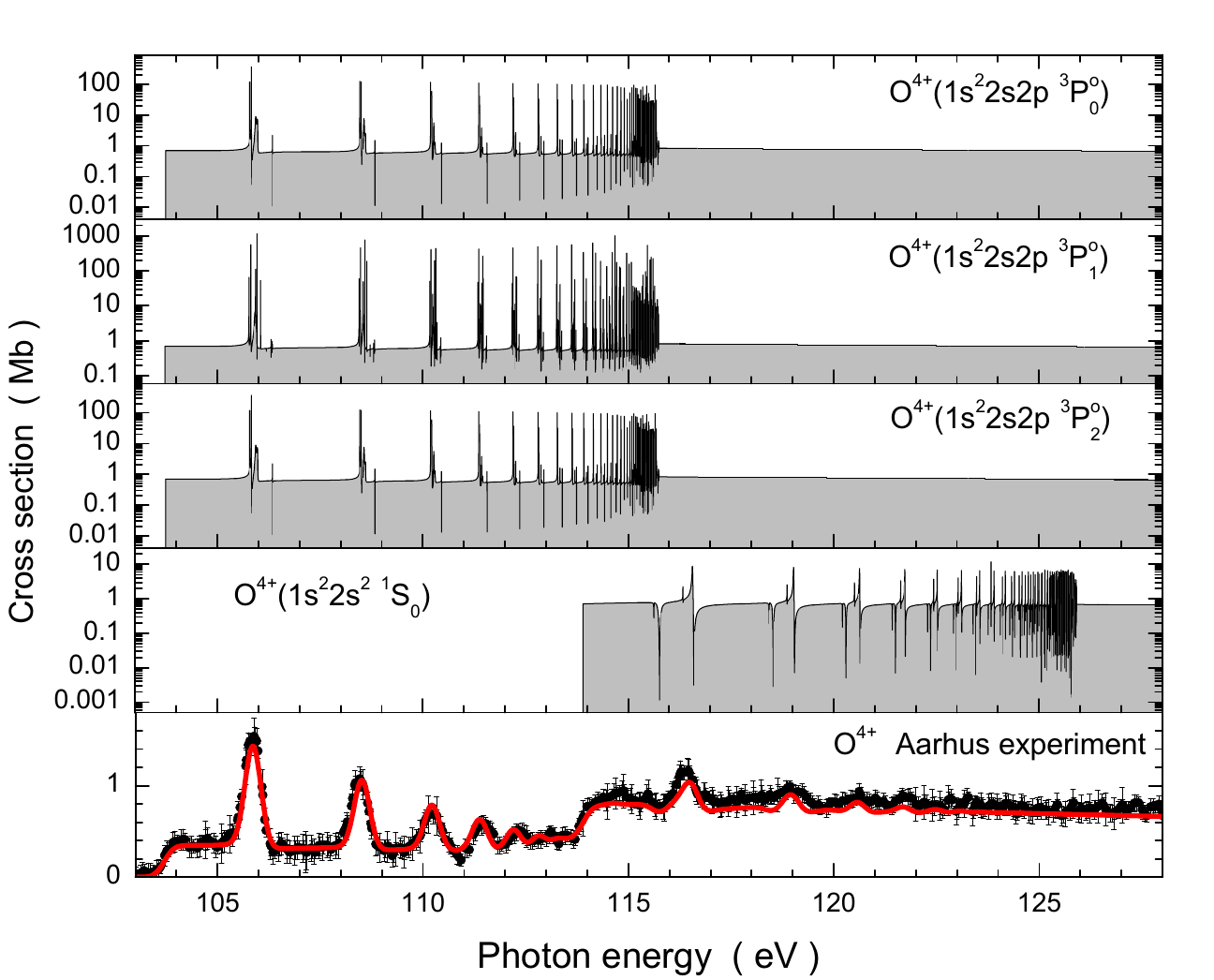}
\caption{\label{fig:O4theoryoverview} (Colour online) Results of the present Breit-Pauli R-matrix calculations for photoionization of O$^{4+}$ ions in the 4 different initial states (1s$^2$\,2s2p~$^3$P$^o_0$) (top panel), (1s$^2$\,2s2p~$^3$P$^o_1$) (second panel from top), (1s$^2$\,2s2p~$^3$P$^o_2$) (third panel from top) and  (1s$^2$\,2s$^2$~$^1$S$_0$) (fourth panel from top) together with an overview experiment (bottom panel) by Bizau \etal~\cite{Bizau2005a}. The experimental data are shown as open circles with statistical error bars. The solid (red) line in the bottom panel is a simulation of the experiment on the basis of the present calculations assuming an energy spread of 436~meV, a metastable fraction $f_{\rm m}=0.5$ and statistical population of the $^3$P$^o$ fine structure states.   Note the logarithmic scale in the upper 4  panels.
}
\end{center}
\end{figure}

Figure~\ref{fig:O4theoryoverview} shows the results of the present Breit-Pauli R-matrix calculations for photoionization of O$^{4+}$ ions in the 4 different (1s$^2$\,2s$^2$~$^1$S$_0$), (1s$^2$\,2s2p~$^3$P$^o_0$), (1s$^2$\,2s2p~$^3$P$^o_1$) and (1s$^2$\,2s2p~$^3$P$^o_2$) initial states contributing to the experimental cross section. The theoretical features can be recognized in the experimental data displayed in the bottom panel of the figure. These data are from the measurements by Bizau \etal~\cite{Bizau2005a} carried out at the Aarhus photon-ion merged beams facility \cite{Kjeldsen2006a}.  The energy spread of this experiment was strongly underestimated to be about 250~meV.  Comparison of the present theoretical results with the Aarhus experiment indicates an energy spread of 436~meV. Apart from that the agreement of the present calculations and the absolute measurement by Bizau \etal~\cite{Bizau2005a} is excellent. An overall scaling factor of 1.13 applied to the present theory, i.e. a 13\% difference, would further improve the agreement. Discrepancies of this size are  within the error bars of $\pm$15\% estimated by Bizau \etal for their experiment.

\begin{figure}
\begin{center}
\includegraphics[width=\textwidth]{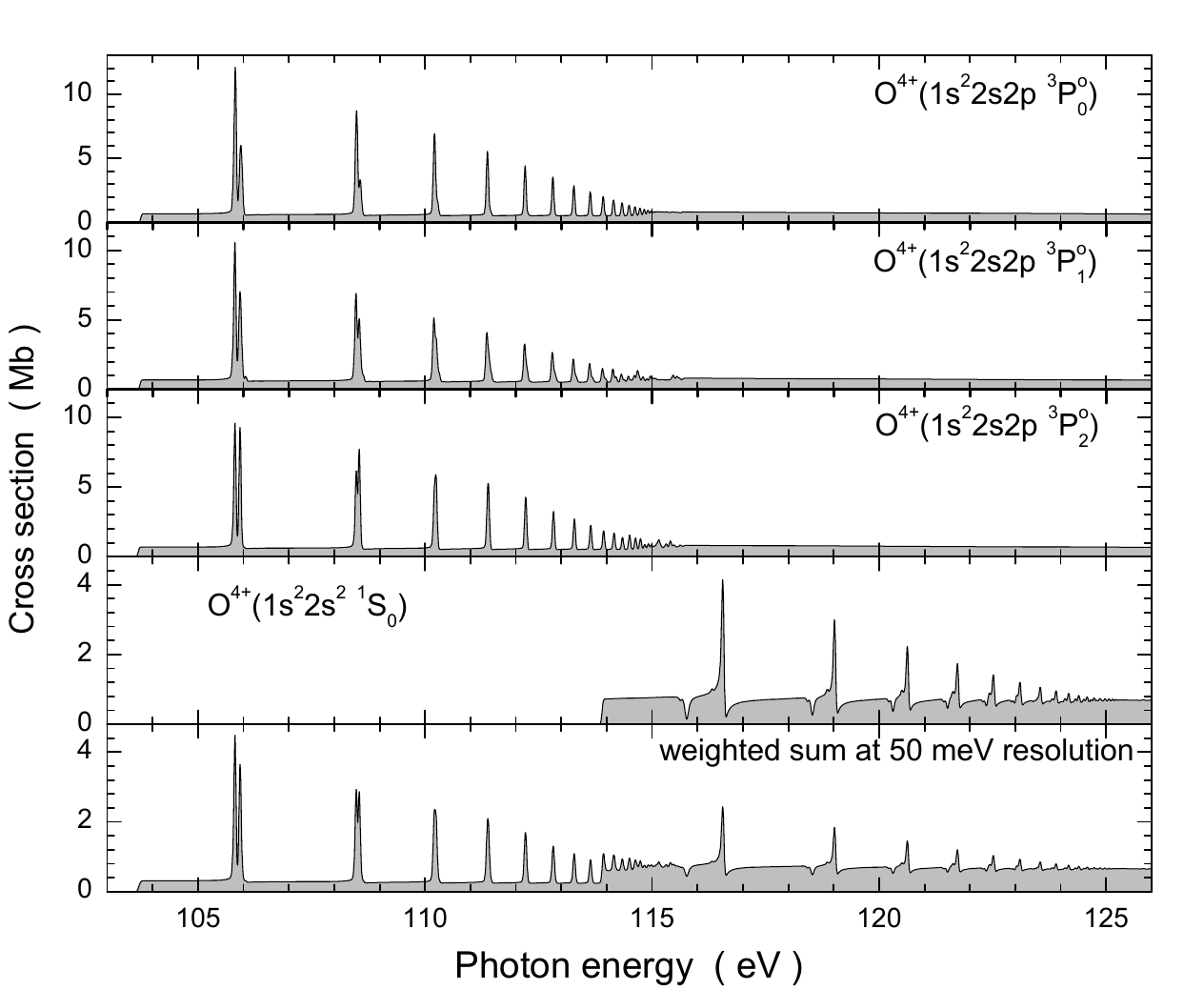}
\caption{\label{fig:O4theoryoverviewconvoluted} (Colour online) Theoretical overview on photoionization of O$^{4+}$ ions with the present Breit-Pauli R-matrix calculations from figure~\ref{fig:O4theoryoverview}  convoluted with a 50~meV FWHM gaussian. The  theoretical model spectrum displayed in the bottom panel simulating an experiment at 50~meV resolution  was obtained by assuming a metastable fraction $f_{\rm m}=0.5$ in the parent ion beam and statistical population of the (1s$^2$\,2s2p~$^3$P$^o_0$), (1s$^2$\,2s2p~$^3$P$^o_1$) and (1s$^2$\,2s2p~$^3$P$^o_2$) metastable states.
}
\end{center}
\end{figure}

 To elucidate the strengths of the individual O$^{4+}$ photoionization contributions from the many resonances calculated within the present Breit-Pauli R-matrix approach for the initial $^1$S and $^3$P$^o$ states, figure~\ref{fig:O4theoryoverviewconvoluted} shows the theoretical data from the first 4 panels of figure~\ref{fig:O4theoryoverview}, this time convoluted with a 50~meV FWHM gaussian. As in the previous examples, the convolution greatly simplifies the photoionization spectra and clearly demonstrates where the dominant resonance strengths are located. The bottom panel simulates an experiment at 50~meV resolution. The spectrum  was obtained by assuming a metastable fraction $f_{\rm m}=0.5$ in the parent ion beam and statistical population of the (1s$^2$\,2s2p~$^3$P$^o_0$), (1s$^2$\,2s2p~$^3$P$^o_1$) and (1s$^2$\,2s2p~$^3$P$^o_2$) metastable states.

\begin{figure}
\begin{center}
\includegraphics[width=\textwidth]{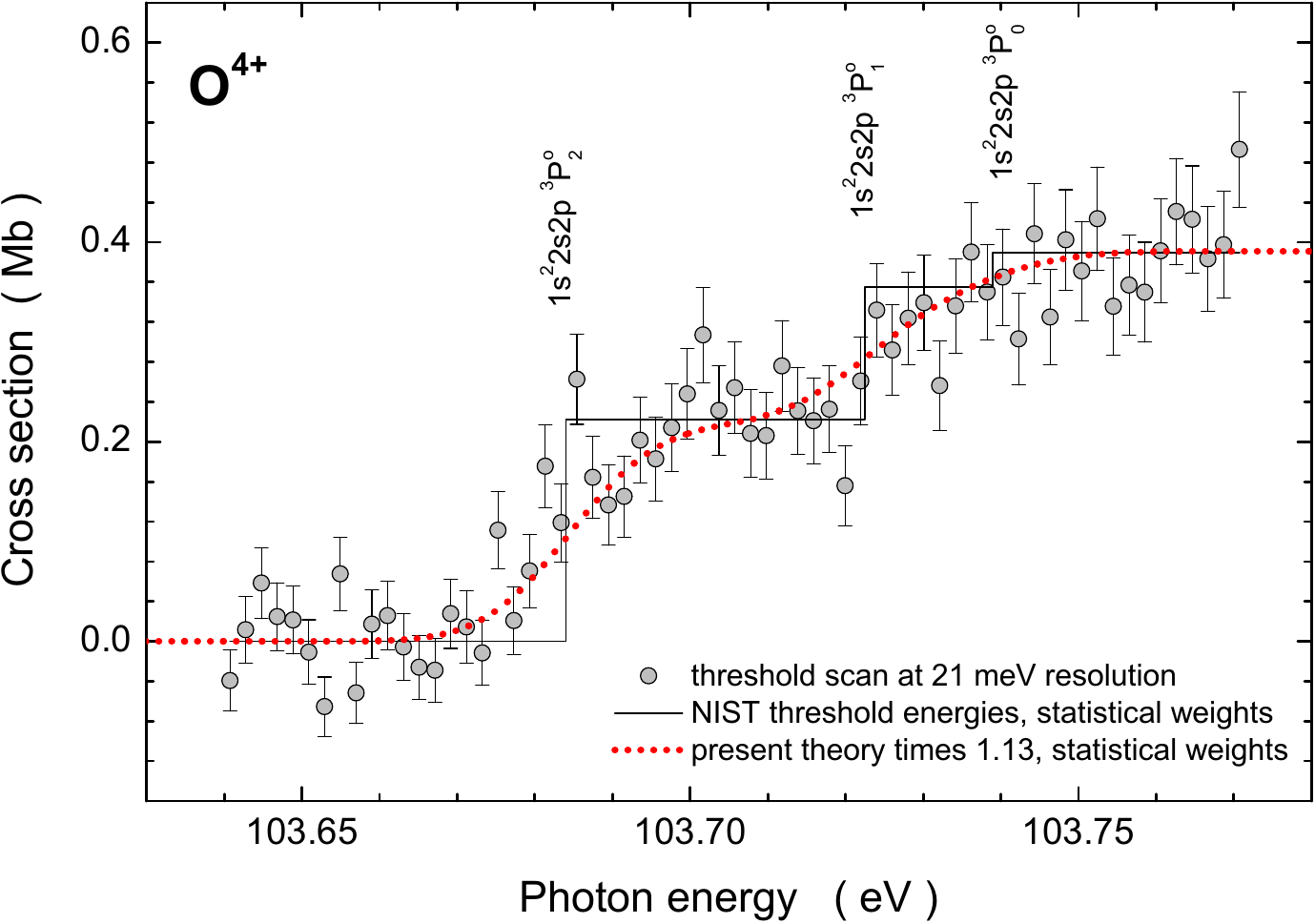}
\caption{\label{fig:O4threshold} (Colour online) Threshold scan for photoionization of O$^{4+}$ ions. The relative cross sections were obtained by a high resolution relative scan measurement and normalized to the absolute measurements of Bizau \etal~\cite{Bizau2005a}. The experimental data are represented by open circles with gray shading. Only statistical error bars are shown. The step function (solid black line) was constructed by using the ionization thresholds \cite{NIST2004} of the individual $^3$P$^o$ fine structure states and distributing the height of the threshold step (0.39~Mb) over contributions of the fine structure states assuming statistical population. The dotted (red) line is from a convolution of the theoretical result with a 21~meV FWHM gaussian. For the theoretical simulation a metastable fraction $f_{\rm m}=0.5$ and statistical population of the fine structure states was assumed. Finally, the theoretical cross sections have been scaled up by a factor 1.13.
}
\end{center}
\end{figure}

The present experimental study was focused on the first 2.5~eV energy range covering the photoionization threshold of O$^{4+}$ and the first group of resonances associated with (2s2p) $\to$ (2p6p) excitations. A threshold scan of the O$^{4+}$ photoionization  at high resolution yielded relative cross sections. These were normalized to the absolute measurements of Bizau \etal~\cite{Bizau2005a}. The resulting normalized high resolution data are shown in  figure~\ref{fig:O4threshold}.  The step function (solid black line) was constructed by using the ionization thresholds \cite{NIST2004} of the individual $^3$P$^o$ fine structure states and distributing the height of the threshold step (0.39~Mb) over contributions of the fine structure states assuming statistical population. The dotted (red) line is from a convolution of the theoretical result with a 21~meV FWHM gaussian which provides excellent agreement with the experiment. For the theoretical simulation a metastable fraction $f_{\rm m}=0.5$ and statistical population of the fine structure states was assumed. Finally, the theoretical cross sections have been scaled up by a factor 1.13 as already discussed in the context of figure~\ref{fig:O4theoryoverview}.

\begin{figure}
\begin{center}
\includegraphics[width=\textwidth]{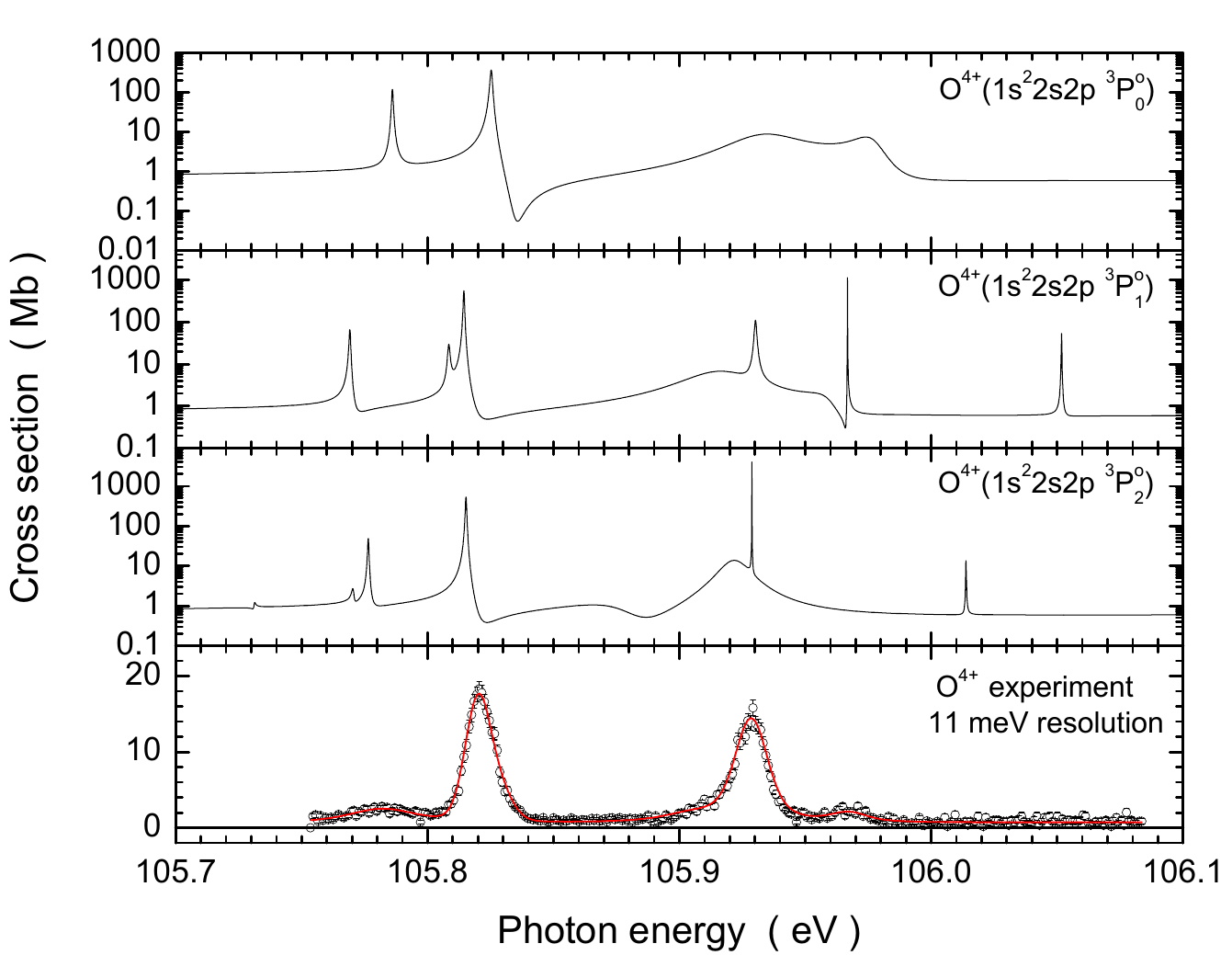}
\caption{\label{fig:O4first2} (Colour online) Results of the present R-matrix calculations in the energy range 105.7~to~106.1~eV for the $^3$P$^o$ initial states of the O$^{4+}$ ion with total angular momenta J=0, J=1 and J=2 (first 3 panels from top). The bottom panel displays the data (open circles with statistical error bars) for a high-resolution photoionization scan in that energy range. The solid (red) line is a fit to the data assuming the presence of 6 contributing resonances in the experimental spectrum. This fit suggested an energy spread of about 11.4~meV. Detailed comparison with the present theory reveal an experimental energy resolution of 11.0~meV in this experiment. Note the logarithmic scale in the upper three panels.
}
\end{center}
\end{figure}

\begin{figure}
\begin{center}
\includegraphics[width=\textwidth]{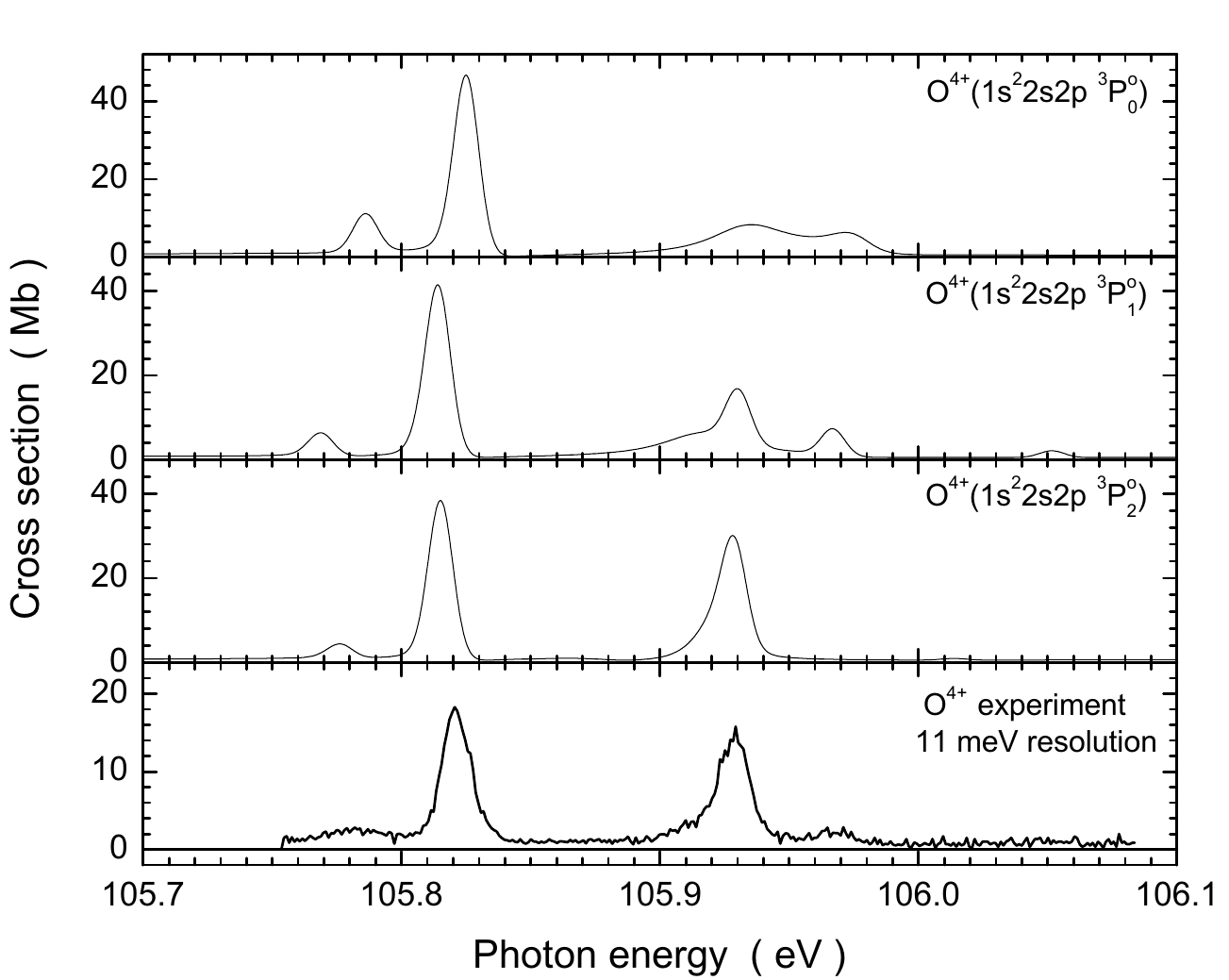}
\caption{\label{fig:O4first2convoluted}  Simulated results of  experiments on photoionization of individual O$^{4+}$(2s2p~$^3$P$^o$) ions with 11~meV resolution in the energy range of figure~\ref{fig:O4first2}. The cross sections in all four panels are on a linear scale. The bottom panel shows the same experimental data as in figure~\ref{fig:O4first2}, this time as a solid line.
}
\end{center}
\end{figure}

As in the cases of the ions C$^{2+}$ and N$^{3+}$ the region of the first two groups of photoionization resonances were experimentally investigated for O$^{4+}$ at the highest possible resolution that could be achieved under the constraints of the measurements. Figure~\ref{fig:O4first2} shows results of the present R-matrix calculations in the energy range 105.7~to~106.1~eV for the $^3$P$^o_{\rm J}$ initial states of the O$^{4+}$ ion with total angular momenta J=0, J=1, and J=2 (first 3 panels). The theoretical data are to be compared with the high resolution experimental cross section shown in the bottom panel. The solid (red) line in the bottom panel is a fit to the data assuming the presence of 6 contributing resonances in the experimental spectrum. The strongest peaks with their asymmetric shapes are represented by two resonances each. This fit suggested an energy spread of about 11.4~meV. Detailed comparison with the present theory provides a better result  for the experimental energy resolution. The spread is found to be 11.0~meV in this experiment. For the detailed direct comparison with experiment the theoretical contributions have to be convoluted with 11.0~meV FWHM gaussians simulating the energy spread in the measurement. The results of such convolution are provided in the first 3 panels of figure~\ref{fig:O4first2convoluted}. The bottom panel shows the experimental data as a solid line.

\begin{figure}
\begin{center}
\includegraphics[width=\textwidth]{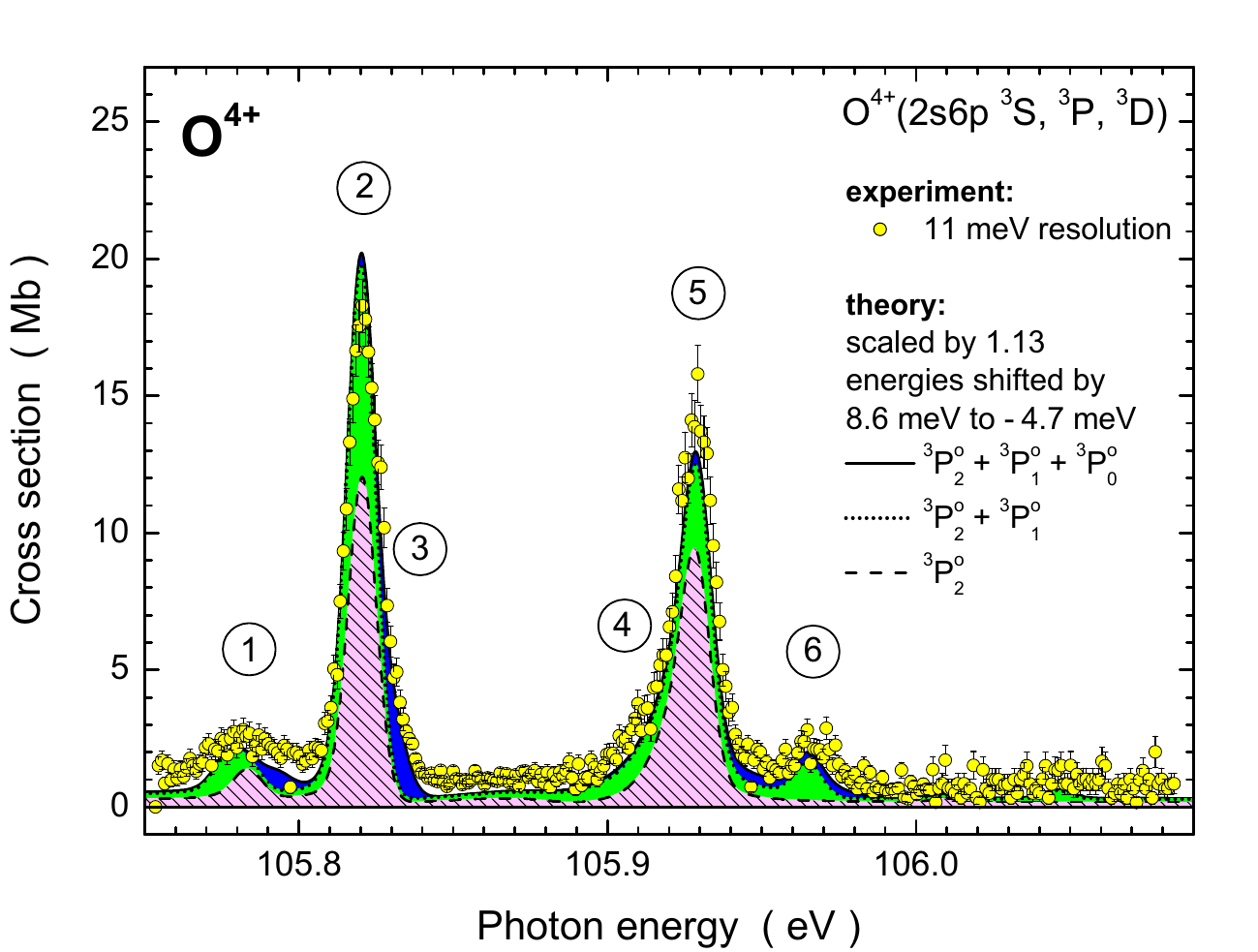}
\caption{\label{fig:O4first2comparison} (Colour online) Illustration of the individual contributions of ions in the (1s$^2$\,2s2p~$^3$P$^o_0$), (1s$^2$\,2s2p~$^3$P$^o_1$) and (1s$^2$\,2s2p~$^3$P$^o_2$) metastable O$^{4+}$ states to the experimental high-resolution scan measurement at a resolution of 11~meV shown already in the previous figures. The scan data are displayed as open (yellow) shaded circles with statistical error bars. The total metastable fraction in the beam was set to $f_{\rm m}=0.5$ in the theoretical simulation. The 5/9 contribution of the $^3$P$^o_2$ fine structure state is shown by the dashed line (hatched area with red shading). On top of that is the 3/9  $^3$P$^o_1$ contribution represented by the difference between the dotted and the dashed line (the green shaded area). Further adding of the 1/9  $^3$P$^o_0$ contribution (the blue shaded area) results in the solid-line envelope of all theoretical contributions to the simulated cross section. The energy scale of the theoretical results was shifted by a linear function of energy resulting in a correction at the low-energy side of the theoretical spectrum by 8.6~meV and -4.7~meV at the high-energy side. Finally, the theoretical cross sections have been scaled up by a factor 1.13. The peak numbers refer to the 6 peaks assumed in the fit to the data as shown in figure~\ref{fig:O4first2}. Peak assignments are provided in table~\ref{tab:LANLmodelO4}. Autoionizing (2p6p~$^3$S,$^3$P,$^3$D) states are formed in the present energy range.
}
\end{center}
\end{figure}

The next step is the modelling of the experiment by calculating the observed apparent cross section as a weighted sum of individual contributions from the different parent ion beam components. As in the analysis of the threshold energy scan (Figure~\ref{fig:O4threshold}), the metastable fraction $f_{\rm m}=0.5$ was assumed. The weights are 1/9 for the (1s$^2$\,2s2p~$^3$P$^o_0$) contribution, 3/9 for the (1s$^2$\,2s2p~$^3$P$^o_1$) contribution, and 5/9 for the (1s$^2$\,2s2p~$^3$P$^o_2$) as determined for C$^{2+}$ and N$^{3+}$. In figure~\ref{fig:O4first2comparison} the experimental data are compared to the theoretical model spectrum obtained by this procedure. In order to provide optimum conditions for detailed comparison, the energy scale of the theoretical results was shifted by adding a linear energy correction function. This shift is 8.6~meV at the low-energy end of the displayed theoretical spectrum and -4.7~meV at the high-energy end. Finally, the theoretical cross sections have been scaled up by a factor 1.13 as discussed above. The figure distinguishes between the contributions of individual initial metastable components in the parent ion beam. Four distinct peak features can be seen at the present resolving power of slightly more than 9,600. The four peaks are composed of different resonance contributions originating from the three different initial $^3$P$^o$ metastable states.

As part of the present analysis, calculations of excitation energies and oscillator strengths relevant to the present region of interest were carried out using the CATS code \cite{LANL,Cowan1981}. From the calculated data shown in table~\ref{tab:LANLmodelO4} one can infer that transitions from metastable initial states (2s2p $^3$P$^o$) to (2p6p $^3$D,$^3$P) excited states are mainly responsible for the features observed in the present energy region of interest. Again the lower-energy structures are mainly due to (2p6p $^3$D) excited states but now also $^1$P$_1$ resonances are calculated to contribute in this energy range with non-negligible strengths.  The higher-energy features originate from (2p6p $^3$P) excited states. The $^3$S levels present in the calculated cross sections are not obvious in the experiment, probably because of the large widths and reduced resonance strengths of these states. Therefore, the three calculated $^3$S contributions are omitted from table~\ref{tab:LANLmodelO4}. Assignments of the resonance features in O$^{4+}$ ions displayed in figure~\ref{fig:O4first2comparison} are inferred from the CATS calculations presented in table~\ref{tab:LANLmodelO4}. The ratios of oscillator strengths $gf$ indicate the relative individual contributions of the specified transitions. Note the shifts of the theoretical data required to match the experiment.

%
%
%

\begin{table}
\caption{\label{tab:LANLmodelO4} Assignments of the resonance features in O$^{4+}$ ions displayed in figure~\ref{fig:O4first2comparison} inferred from calculations employing the CATS code \cite{LANL,Cowan1981}. The ratios of oscillator strengths $gf$ indicate the relative individual contributions of the specified transitions. Note the shifts of the theoretical data required to match the experiment.}
\begin{indented}
 \lineup
 \item[]\begin{tabular}{ccr@{\,}c@{\,}llcl}
\br
 peak    & energy & \multicolumn{3}{c}{transition}   & \multicolumn{1}{c}{energy} & \multicolumn{2}{c}{$gf$}\\
 no.      & (exp., eV) & \multicolumn{3}{c}{(2s2p $\to$ 2p6p)} & \multicolumn{1}{c}{(CATS, eV)} & \multicolumn{2}{c}{(CATS)}\\
 \ns
 \mr
 1 & 105.782   	& $^3$P$^o_1$     &$\to$  	& $^1$P$_1$    & 105.776$^a$	& 0.00321\\
 &             	& $^3$P$^o_2$     &$\to$  	& $^3$D$_2$    & 105.785$^a$	& 0.00384\\
 &             	& $^3$P$^o_0$     &$\to$  	& $^1$P$_1$    & 105.794$^a$	& 0.00200\\

2 & 105.820     & $^3$P$^o_1$     &$\to$  	& $^3$D$_1$    & 105.813$^a$   	& 0.00188\\
  &             & $^3$P$^o_2$     &$\to$  	& $^3$D$_3$    & 105.822$^a$   	& 0.03474\\
  &             & $^3$P$^o_1$     &$\to$  	& $^3$D$_2$    & 105.822$^a$   	& 0.02073\\

3 & 105.829     & $^3$P$^o_0$     &$\to$  	& $^3$D$_1$    & 105.831$^a$   	&  0.00819\\

\br

4 & 105.905     & $^3$P$^o_2$     &$\to$  	& $^3$P$_1$    & 105.921$^b$   	&  0.01062\\

5 &  105.929    & $^3$P$^o_1$     &$\to$  	& $^3$P$_0$    & 105.930$^b$   	&  0.00492\\
  &             & $^3$P$^o_2$     &$\to$  	& $^3$P$_2$    & 105.930$^b$   	&  0.02086\\

6 &  105.967    & $^3$P$^o_1$     &$\to$  	& $^3$P$_1$    & 105.957$^b$   	&  0.00194\\
  &             & $^3$P$^o_1$     &$\to$  	& $^3$P$_2$    & 105.966$^b$   	&  0.00360\\
  &             & $^3$P$^o_0$     &$\to$  	& $^3$P$_1$    & 105.976$^b$   	&  0.00230\\

\br
\end{tabular}
~\\
$^{a}$shifted by -0.311~eV\\
$^{b}$shifted by -0.330~eV\\

\end{indented}
\end{table}

\section{Conclusion}

In the present study, photoionization experiments on four-electron Be-like C$^{2+}$, N$^{3+}$ and O$^{4+}$ ions were carried out with ion beams containing not only ions in their (1s$^2$\,2s$^2$~$^1$S$_0$) ground state but also in the metastable (1s$^2$\,2s2p~$^3$P$^o_0$), (1s$^2$\,2s2p~$^3$P$^o_1$), and (1s$^2$\,2s2p~$^3$P$^o_2$) states. Relative cross sections were obtained by scanning the photon energy and observing the ionization signal, the ion beam current and the photon flux. The relative data were normalized to published absolute measurements by M\"{u}ller \etal~\cite{Mueller2002b} and by Bizau \etal \cite{Bizau2005a}. For several prominent features in the photoionization spectra the experimental energy resolution was pushed to the limit of feasibility in order to scrutinize the details of individual transitions between initial and final states and their contributions to the photoionization cross sections. Resolving powers of up to 21,200 were accomplished. The measurements were accompanied by state-of-the-art Breit-Pauli R-matrix calculations showing extremely good agreement with even the finest details revealed in the high resolution experiments. Supporting calculations employing the online version of the CATS Los Alamos National Laboratory Atomic Physics Codes package \cite{LANL,Cowan1981} facilitated assignments of observed structures and resonances to individual transitions  $^3$P$^o_{\rm J} \to ^3$D$_{\rm J'}$ and $^3$P$^o_{\rm J} \to ^3$P$_{\rm J''}$ with J=0,1,2, J'=1,2,3 and J''=0,1,2. The oscillator strengths $gf$ for the most important photoexcitation processes were calculated on the same footing providing additional information about the relative strengths of individual state-to-state selective contributions to the observed resonance features in the photoionization cross sections.

For such a comparison of theory and experiment,  the fractions of the ground state and the long lived excited states have to be assessed. A plausible assumption about the relative fractions of the three initial metastable  $^3$P$^o_{0,1,2}$ states is a statistical population of the fine structure states, i.e., relative weights of 1/9 : 3/9 : 5/9  for total angular momenta J=0,1,2, respectively. The ratio of metastable ($f_{\rm m}$) versus ground state ($f_{\rm g}$) ion components in the primary ion beam was found to be about $f_{\rm m}$ : $f_{\rm g}$ = 0.5 : 0.5 with possible excursions of that ratio from 0.4 : 0.6 to 0.6 : 0.4. While the detailed comparisons of theory and experiment always support  statistical population within the $^3$P$^o_{0,1,2}$ manifold, the fractions $f_{\rm m}$ could not be determined equally well because of the uncertainties of the absolute experimental cross sections. Considering the related limitations of the experimental data, the agreement between theory and the high resolution experiments even in finest details of the cross section functions is truely remarkable. Evidently, the present   state-of-the-art Breit-Pauli R-matrix approach is very well suitable for providing excellent predictions of photoionization cross sections for Be-like ions in low charge
states even at the level of state-to-state resolved data.

\ack
We acknowledge support by
Deutsche Forschungsgemeinschaft under project number Mu 1068/10  and through
NATO Collaborative Linkage grant 976362 as well as by the US Department of Energy (DOE)
under contract DE-AC03-76SF-00098 and grant  DE-FG02-03ER15424.
B M McLaughlin acknowledges support by the US
National Science Foundation through a grant to ITAMP
at the Harvard-Smithsonian Center for Astrophysics.
The computational work was carried out at the National Energy Research Scientific
Computing Center in Oakland, CA, USA and on the Tera-grid at
the National Institute for Computational Science (NICS) in TN, USA,
which is supported in part by the US National Science Foundation.

%
%
%
%
~\\~\\

	

\end{document}